\documentclass[preprint]{aastex} 

\usepackage{color}
\usepackage{graphicx}
\usepackage{epsfig}
\usepackage{natbib}
\usepackage{rotating}
\usepackage{amsmath}

\begin{document}

\title{The Coldest Brown Dwarf (Or Free Floating Planet)?: The Y Dwarf WISE 1828+2650}
\author{C. Beichman$^{1,2,3}$, Christopher R.\ Gelino$^{1,3}$, J.\ Davy Kirkpatrick$^1$, 
Travis S.\ Barman$^4$, Kenneth A.\ Marsh$^{1,5}$, Michael C.\ Cushing$^6$, and E.L. Wright$^7$}
\affil{1) Infrared Processing and Analysis Center, California Institute of Technology, Pasadena CA 91125}
\affil{2) Jet Propulsion Laboratory, California Institute of Technology, 4800 Oak Grove Dr., Pasadena CA 91107}
\affil{3) NASA Exoplanet Science Institute, California Institute of Technology,  770 S. Wilson Ave., Pasadena, CA 91125}
\affil{4) Lowell Observatory, 1400 W.\ Mars Hill Road, Flagstaff, AZ 86001}
\affil{5) School of Physics and Astronomy, Cardiff University, Cardiff, Wales, CF24 3AA }
\affil{6) Department of Physics and Astronomy, The University of Toledo,
  2801 West Bancroft Street, Toledo, OH 43606}
\affil{7) UCLA Physics \& Astronomy, PO Box 951547, Los Angeles CA 90095}

\email{chas *at* ipac.caltech.edu}

\begin{abstract}

We have monitored the position of the cool Y dwarf WISEPA J182831.08+265037.8 using a combination of ground- and space-based telescopes and have determined its distance to be 11.2$_{-1.0}^{+1.3}$ pc. Its absolute H magnitude, M$_H=22.21^{+0.25}_{-0.22}$ mag, suggests a mass in the range 0.5-20 M$_{Jup}$ for ages of  0.1-10 Gyr  with an effective temperature in the range 250-400 K. The broad range in mass is due primarily to the unknown age of the object. Since the high tangential velocity of the object, 51$\pm5$ km s$^{-1}$, is characteristic of an old disk population, a plausible age range of 2-4  Gyr leads to a mass range of 3-6  M$_{Jup}$ based on fits to the (highly uncertain) COND evolutionary models. The  range in temperature is due to the fact that no single model adequately represents the 1-5 $\mu$m spectral energy distribution (SED) of the source, failing by factors of up to 5 at either the short or long wavelength portions of the spectral energy distribution. The appearance of this very cold object may be affected by non-equilibrium chemistry or low temperature condensates forming clouds, two atmospheric processes that are known to be important in brown dwarf atmospheres but have proven difficult to model. Finally, we argue that there would have to be a very steep upturn in the number density of late type Y-dwarfs to account for the putative population of objects suggested by by recent microlensing observations. Whether WISE 1828+2650 sits at the low mass end of the brown dwarf population or is the first example of a large number of ``free-floating" planets is not yet known.

\end{abstract}

\keywords{brown dwarfs – infrared:stars - individual (WISEP J182831.08+265037.8) -  Astrometry - 
   Parallaxes - proper motions - solar neighborhood}

\section{Introduction}


A long term goal of infrared sky surveys has been the detection and characterization of sub-stellar objects of progressively  lower temperature and mass \citep{Kirkpatrick1997}. The Two Micron All sky survey (2MASS), DEep Near Infrared Southern Sky Survey (DENIS) and Sloan Digital Sky Survey \citep{Skrutskie06,Epchtein97,York2000} identified the first bona fide field brown dwarfs while coronagraphic studies found the first bona fide brown dwarf companion, Gliese 229B \citep{Nakajima1995, Oppenheimer1998}. These studies led to the acceptance of the first new spectral classes to be named in over 50 years, the L and T dwarfs \citep{Kirkpatrick1999} characterized by effective temperatures, $T_{\rm eff}$, as low as 600 K. Studies of these objects  suggested the potential for yet a third class of even colder objects, the Y dwarfs, with temperatures between 300-600 K. Many examples of L and T dwarfs have now been discovered and their spectral classification is now a fine art \citep{davy2005}. But members of the Y spectral class were elusive until the launch of the Wide-field Infrared Survey Explorer (WISE; Wright et al 2010) which identified over 100 brown dwarfs with low effective temperature, including the   first Y dwarfs. \citep{Kirkpatrick2011, Cushing2011}. 

In parallel with the WISE survey, targeted imaging of stars has led to the discoveries of planetary mass objects orbiting nearby stars, yielding both  young planets, e.g. HR8799 bcde \citep{Marois2008}, $\beta$ Pic b \citep{lagrange} and 2M1207-3932b \citep{chauvin}, as well as  a companion to a  nearby white dwarf, WD 0806-661 b \citep{Luhman2011, Luhman2012}. Thermal emission has also been detected from Jovian mass planets transiting their parent stars \citep{Deming2006, Charbonneau2005}. Finally, microlensing studies may have  identified a population of Jovian-mass objects with no obvious association to any star and which thus must either be located tens of AU from their hosts or be free floating planets \citep{Sumi2011}. 

Essential to understanding the nature, physical properties and evolutionary status of these objects is the determination of such properties such as mass, age, composition, and bolometric luminosity. This article describes the first results of a program to use direct imaging to determine distances to these ultra-cool brown dwarfs using parallax measurements. With secure distance measurements it will be possible to estimate the local space density of Y dwarfs and to understand better their physical properties.
 
We have taken a sample of twenty objects from the entire all-sky WISE survey that are confirmed to have spectral
types (or types estimated from WISE photometry)  of T8 and later and have initiated a program of follow-up photometry from ground- and space-based telescopes. From a comparison of apparent and inferred absolute magnitudes for objects of low T$_{\rm eff}$, e.g.  \citet{Baraffe2003, Burrows2003}, we estimate that these objects are within 20 pc of the Sun and so shoud have large astrometric signatures. We are using the {Hubble Space Telescope} ({\it HST}), the {\it Spitzer Space Telescope}, and the Keck-II telescope to obtain images with the goal of determining parallactic distances. This paper focuses on the first result of this program, the determination of the distance to WISEPA J182831.08+265037.8, hereafter, WISE 1828+2650 which was identified as a potential late-type brown dwarf on the basis of its extreme color, W1-W2 ([3.5\micron]-[4.6\micron]) $>$ 4 mag \citep{Kirkpatrick2011}. Subsequent near-infrared observations from {\it HST} led to its classification as the prototype  Y dwarf \citep{Cushing2011}, and finally, on the basis of a broader sample of Y dwarf discoveries have revised the classification to $\ge$Y2 \citep{Kirkpatrick2012}.

\section{Observations}

The determination of high accuracy positions for cool, low luminosity brown dwarfs is extremely challenging since they are faint  for 8-10 m ground-based telescopes and  HST in the near-infrared and bright in the mid-IR where they can only be observed with space telescopes of small aperture (0.4-0.85 m). As a result we have had to piece together positional information with a variety of telescopes over a range of wavelengths using individual measurements (Table~\ref{ObsLog}) with uncertainties in the range of 10-50 mas due to either low signal to noise ratio or large beam sizes. A  parallax determination  will be  possible only if these objects prove to quite close-by. Fortunately, this appears to be the case for at least WISE 1828+2650 where we are able to determine  a parallax of $\sim$100 mas.  With such a large astrometric signature, relatively  coarse positions  suffice and many of the instrumental effects that must be taken into account for more distant objects can be ignored. Table~\ref{ObsLog} lists the observations from WISE and other observatories. Photometric data are presented in Table~\ref{phot}. For WISE we report an average  of the 4.6 \micron\ magnitudes from the two epochs and $3\sigma$ upper limits in the other bands. For comparison with models we have converted the magnitudes to flux densities using  the zero points from  \citet{Wright2010}, but because of the unknown and extremely non-blackbody-like nature of brown dwarf spectral energy distributions (SED), we have not color-corrected these flux densities.

\subsection{HST Observations}
\begin{figure*}
\includegraphics[scale=0.75]{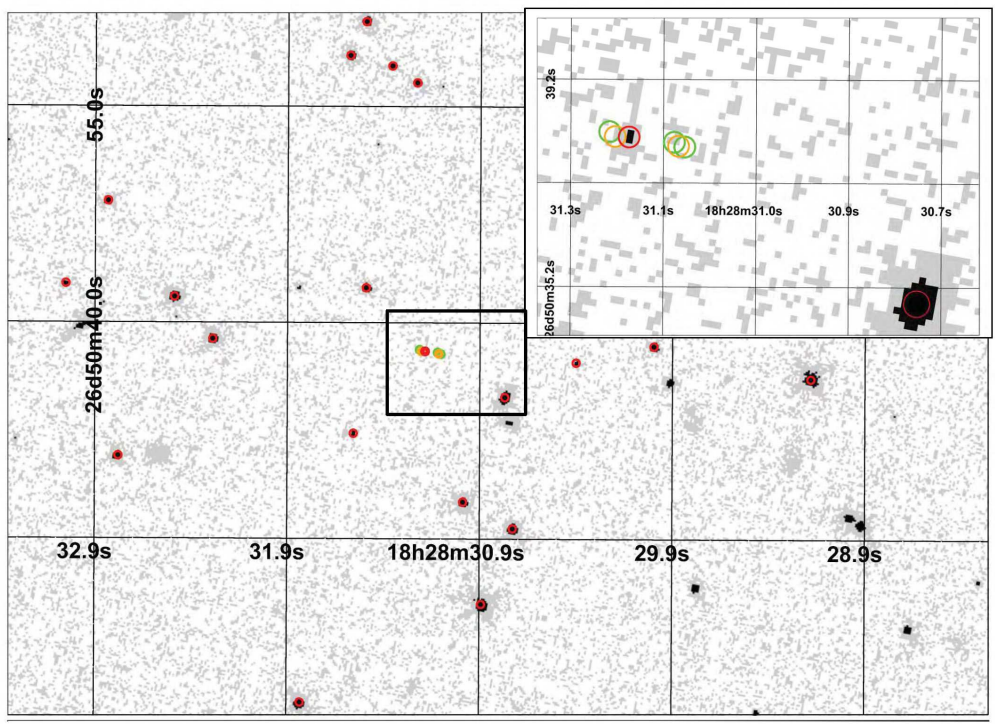} \caption{An {\it HST} WFC3/F140W image of WISE 1828+2650  with the reference stars used for the co-registration of the {\it Spitzer}, {\it HST}, and Keck fields circled in red; a few of the 2MASS reference stars fall outside of the image shown here to preserve an adequate scale in the vicinity of the WISE source. The absolute positions were derived with respect to 11 2MASS stars spread over the entire {\it HST} field. The positions of WISE 1828+2650  at different epochs are shown in an inset with {\it HST} positions in red, {\it Spitzer} positions in green, and Keck positions in orange. The alternate astrometric reference denoted ``Source A" appears to be slightly extended and is located to the southwest of WISE 1828+2650.\label{HST}}
\end{figure*}

WISE 1828+2650 was observed with {\it HST} in the WFC3-IR F140W filter in conjunction with a grism spectrum which was the basis of this object's identification as a $\ge$Y1 dwarf by \cite{Cushing2011} (Figure~\ref{HST}). The data consist of two pairs of individual 78 sec exposures  taken at positions offset by ($\Delta\alpha,\Delta\delta)= $0.82\arcsec\ and 5.67\arcsec\ to  reduce the effects of cosmic rays and  the  undersampling of the individual frames. The pipeline  corrects for the geometric distortion of the WFC3-IR camera to a level estimated to be $\sim$ 5 mas \citep{Kozhurina2009} which is of the same order or less than  the extraction uncertainties of the faint target. Over 200 sources were extracted from  the  Space Telescope Science Institute's (STScI) ``MultiDrizzle"  mosaic  using the Gaussian-fitting IDL FIND routine to determine centroid positions and  the APER routine\footnote{All the photometric measurements reported herein were made using this routine from the IDL ASTRO library, http://idlastro.gsfc.nasa.gov/contents.html. A number of other IDL routines are taken from this library as well.} with a 3 pixel radius for photometric measurements. The Full Width at Half Maximum (FWHM) for these undersampled data is 0.26\arcsec\  consistent with STScI analyses \citep{Kozhurina2009}. The data  were calibrated using the zero-point for Vega magnitudes, ZP(Vega)=25.1845 mag\footnote{http://www.stsci.edu/hst/wfc3/phot\_zp\_lbn}, from the WFC3 Handbook \citep{Rajan2010}.  A strong source denoted Source A appears to be slightly extended and is located roughly 5\arcsec\ W and 3\arcsec\ S of WISE 1828+2650. This object is seen in the Keck, {\it HST}, and {\it Spitzer} images and will be used as a local astrometric reference in subsequent analysis. As discussed below, the extracted position of Source A is consistent between the Keck and HST measurements to within 10 mas which is small compared to the other uncertainties.

\subsection{Keck NIRC2 Observations}

WISE 1828+2650 was observed in the H-band using the Keck-2 telescope (July 2010) using NIRC2 with the laser guide star Adaptive Optics (AO) system \citep{Wizinowich2006,vanDam2006} and a tip-tilt star located 50\arcsec\ away.  The wide-field camera (40 mas/pixel scale; 40\arcsec\ field of view) was used  to ensure an adequate number of reference stars for astrometry. A sequence of images with a total integration time of  1080 sec was obtained at airmass of 1.0-1.1 with 2\arcsec--10\arcsec\ offsets in right ascension and declination \citep{Kirkpatrick2011}. The individual images were  sky-subtracted with a sky frame created by the median of the science frames and flat-fielded with a dome flat using standard and custom IDL routines. Individual images were ``de-warped'' to account for optical distortion in the NIRC2 camera as described below. The reduced images were shifted to align stars on a common, larger grid and the median average of overlapping pixels was computed to make the final mosaic. The target was detected at a relatively low SNR of $\sim$5. As reported in \citet{Kirkpatrick2012}, the source was also weakly detected in the J band with NIRC2 at this epoch (Table~\ref{phot}).

\begin{figure*}
\includegraphics[scale=0.75]{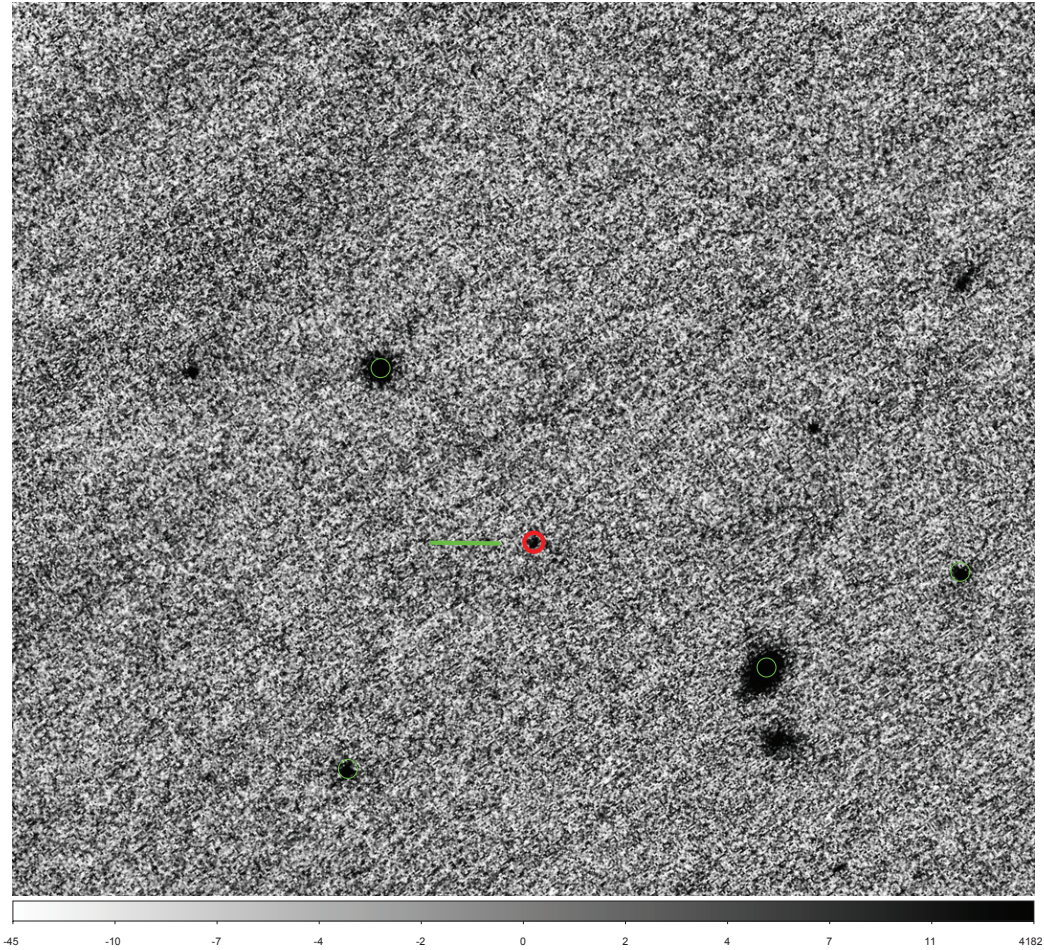} \caption{A {\it Keck-II} H-band image of WISE 1828+2650 obtained with the NIRC2 instrument using the laser guide star Adaptive Optics system. The image is 25\arcsec\ across with North up and East to the left. The solid line is 2\arcsec\ in length and sits just to the East of WISE1828+2650 which is circled in red. Also shown are three of the reference stars used to establish the astrometric reference (Figure~\ref{HST}). The alternate astrometric reference (``Source A") is circled and is located to the southwest of WISE 1828+2650. \label{Keck}}
\end{figure*}

The source was re-observed in 2011 (Figure~\ref{Keck}) and twice in 2012 following the same procedure but for a longer total integration times (3960, 2880 and 2880 sec, respectively), yielding SNR$\sim10$ and with smaller offsets (2-4\arcsec) to reduce the effects of optical distortion on the final image. The airmass for the Oct 2011 observations ranged from 1.1 to 1.4 with a mean value of 1.25. In 2012 July the airmass values ranged between 1.04 to 1.15 with a mean value of 1.09. In 2012 October, the seeing and photometric conditions were relatively poor with an average airmass of 1.2. The typical Point Spread Function (PSF) in the final mosaicked images from the first three epochs varies slightly across the field, but at the center of the field close to WISE 1828+2650, the Full Width Half Maximum (FWHM) of the AO-corrected core is $\sim$3-4 pixels, or $\sim$120-160 mas. In October 2012, the image quality was about a factor of two worse as is reflected in the poorer astrometric solution described below.  Sources were extracted using the IDL APER routine with an aperture radius of 5 pixels. To  calibrate these data photometrically we used observations of the WISE 1828+2650 field obtained with the Wide Field Infrared Camera (WIRC) \citep{Wilson2003} at the Palomar 5-m telescope with a MKO-NIR H filter \citep{Tokunaga2002}. The 8.7\arcmin\ field of this camera was sufficient to detect $\approx$350 2MASS stars surrounding WISE 1828+2650 as well as 6 stars seen in the Keck  field. While the Palomar image was not deep enough to detect WISE1828+2650 itself, we were able to calibrate the Keck field using the calibrated WIRC magnitudes of the 6 stars in common. Combined with the 2MASS calibration uncertainties \citep{Carpenter2001}, the overall calibration uncertainty is $\sim$5\% which is small compared to the uncertainty in the average magnitude for the WISE target itself ($\sim9$\%). There is no evidence for variability between the different epochs. No photometry was attempted for 2012 October.

The source positions obtained from the Keck images were corrected for the effects of differential refraction relative to the center of the field using meteorological conditions available at the CFHT telescope weather archive\footnote{http://mkwc.ifa.hawaii.edu/archive/wx/cfht/} to determine the index of refraction corrected for wavelength, local temperature, atmospheric pressure and relative humidity \citep{Lang1983} and standard formulae \citep{Stone1996}. As discussed in Appendix I, for the small field of view of the NIRC2 images and  the relatively low airmasses under consideration here ($<$1.5), the first order differential corrections we do apply to the source positions are small, $<$10 mas across the $\pm$20\arcsec\ field and proportionately less at smaller separations.

The effects of optical distortion in the wide-field NIRC2 camera were corrected using a distortion map derived by comparing Keck data of the globular cluster M15 taken immediately after the WISE 1828+2650 observations in 2011 and July 2012 to HST/WFC3 observations of M15 \citep{Anderson2008,Yelda2010}. Details of this distortion mapping are described in Appendix I but the correction amounts to $<1$ pixel (40 mas) across most of the array and up to 2 pixels at the edges of the array. At the array center where the WISE 1828+2650 observations were made, the maximum uncorrected distortion for separations up to 5\arcsec\ is $<$ 20 mas. After our correction procedure the residual distortion errors are less than 10 mas over the entire field. As a test  of our ability to co-register data from individual Keck images, we examined the positional dispersions  of 15 sources in each apparition with respect to their  average positions. For the first three Keck epochs the average dispersion in right ascension and declination is $\pm$6 mas. The last epoch was not considered in this analysis due to poor observing conditions. The ability to register these images to an external reference, i.e. the sources in the HST image, is discussed below ($\S$\ref{astromsection}).

\subsection{Spitzer Observations}

{\it The Spitzer Space Telescope} observed WISE 1828+2650 using Director's Discretionary Time and at three subsequent 
General Observer epochs. The IRAC camera \citep{Fazio2004} was used in its  full array mode to make observations at 3.5 (Channel-1) and 4.5 $\mu$m (Channel-2) in the first  epoch and only at 4.5 $\mu$m in the subsequent  epochs. We analyzed post-BCD mosaics from the Spitzer Science Center (SSC) to make photometric and astrometric measurements, extracting sources using a 3 pixel radius aperture and normalized using SSC-recommended aperture corrections\footnote{http://irsa.ipac.caltech.edu/data/SPITZER/docs/irac}. WISE 1828+2650 was detected at both 3.5  \micron\ (SNR$\sim10$) and at 4.5 \micron\ (SNR$>$40; Table~\ref{phot}). Source A is seen as a weak but definite detection with about one-sixth the brightness of WISE 1828+2650 (SNR=9). We fitted the pixel data for these two objects simultaneously using Gaussian profiles to determine accurately the source centroids. The fitting uncertainties for the two objects in the 4 Spitzer observations are 10-20 mas for WISE 1828+2650 and 120-130 mas for Source A (Table~\ref{astromSrcA}). There was no evidence for variability in the brightness of WISE 1820+2650 between the four epochs at the $<$6\% level. Since Source A appears extended at shorter wavelengths, there is a concern that  position of Source A might shift with wavelength. There is no evidence for this effect in our data. The  Spitzer  position for Source A is self-consistent within 15 mas over 4 epochs and consistent with the Keck and HST positions to within ($\Delta\alpha$, $\Delta\delta$)=(60,10) mas which is small relative to the fitting uncertainties for Source A of 120 mas.

The IRAC camera has significant optical distortions over its entire field that are corrected in the SSC pipeline to an average residual value of 100 mas over the whole field\footnote{http://irsa.ipac.caltech.edu/data/SPITZER/docs/irac/iracinstrumenthandbook/26/}. We have used $\sim$20 HST stars observed within $\pm$90\arcsec\ of WISE 1828+2650 as probes of the residual distortion as described above for the Keck observations.  As a test  of our ability to co-register data from individual Spitzer images, we examined the positional dispersions  of 22 sources in each apparition with respect to their  average positions. The average dispersion in the four Spitzer epochs is $\pm$60 mas in right ascension and declination. The ability to register these images to an external reference, i.e. the sources in the HST image, is discussed below ($\S$\ref{astromsection}). These values are less than the quoted 100 mas uncertainty from the SSC, in part because we are confining our observations to the central part of the IRAC array. The total uncertainties in Table~\ref{astromdata} reflect the systematic uncertainties combined with a brightness-dependent term for WISE1828. In the case where we consider only offsets from WISE 1828+2650 to Source A, the distortion effects will be considerably smaller given the $\sim$10 pixel distance between the two objects. From examination of the polynomial expression to the distortion correction given by the SSC, we estimate the distortion component of the uncertainty in the separation between the two sources to be $<$10 mas, negligible compared to the uncertainties due to the source fitting (Table~\ref{astromSrcA}).

\section{Astrometric Data Reduction \label{astromsection}}

Following source extraction, we put observations from all telescopes onto a common astrometric frame. This was accomplished in a two-step process. First, to ensure consistency with the WISE positions which are based on the 2MASS astrometric frame \citep{Wright2010}, the {\it HST} positions were refined using 11 2MASS sources to solve for the RA and DEC zero points and image rotation. We adopted the nominal STScI pipeline plate scale of 0.12826 arcsec/pixel  which is known to an accuracy $\sim0.01$\%   \citep{Kozhurina2009}. This value agrees with the plate scale derived from the 2MASS sources to within $<$0.02\%, but we adopted the better established pipeline value.  Saturation effects on the astrometric centroid for the faint 2MASS sources (H$>$15 mag)  are small ($<<$ the 0.128\arcsec\ HST pixel) compared to the overall 80 mas uncertainty in the absolute 2MASS reference frame  \citep{Skrutskie06}. Additional evidence that the saturation effects are relatively small comes from the fact that our HST F140W photometry recovers the 2MASS brightness  of these 11 objects to within an average  $\pm$0.2 mag where we compared the HST data with the simple  average of the 2MASS J and H magnitudes. The fit to the {\it HST}/2MASS sources yielded a ($\Delta\alpha\, ,\Delta\delta$)(initial HST-2MASS)=(-0.50$\pm0.02$,0.15$\pm0.05$)\arcsec\  shift relative to the initial pipeline solution where we adopted the positional uncertainties from the 2MASS catalog to weight the individual sources (70-350 mas with an average value of 180 mas) in the fitting procedure. We believe the 2MASS positional zero points to be correct given the agreement between 9 sources within 2\arcmin\  of WISE1828+2650 appearing in both the  2MASS and the UCAC3 catalogs \citep{Zacharias2010}. The average positional differences between these two catalogs is  $(\Delta\alpha\, ,\Delta\delta)$(UCAC3-2MASS)=(0.12$\pm0.01$,0.03$\pm0.02$)\arcsec.

Sources seen in the HST image and seen in either the Keck, Spitzer, or 2MASS data are presented in Table~\ref{HSTposn}.
The positional errors associated with the HST data come from a combination of a brightness-dependent term, $\sigma_{SNR}={\rm FWHM}/(2\times {\rm SNR)}$ \citep{Monet2010} where SNR is the signal-to-noise ratio of a particular measurement, and two terms  term depending on distance of the source, $R$, from the center of the HST field, $\sigma_{\rm rot}=R d\theta$ and $\sigma_{\rm scale}=R d{\rm(scale)/scale}$.  The latter two terms account for uncertainty in the rotation angle and plate scale.  The rotational uncertainty is $d\theta=0.028^o=4.8\times10^{-4}$ rad (Table~\ref{astrometry}). For sources at the edge of the field the positional uncertainty can be as large as 90 mas, but within 20\arcsec\ of the center of the Keck field, for example, the typical uncertainty of an HST source is $\sim$5 mas.

Parameters of the 2MASS/HST fit are given in Table~\ref{astrometry}. As shown in the first entry  in Table~\ref{astrometry}, the standard error ($\sigma/\sqrt{N}$) in this local frame is approximately 4 mas in each axis. To this local registration uncertainty must be added the overall astrometric accuracy of the 2MASS survey of $\sim$80 mas \citep{Skrutskie06}. However for the differential measurements of interest for the parallax determination,  this uncertainty is not  important. The new solution was applied to all the {\it HST} sources, of which 13-22 appear in various of the Keck and {\it Spitzer} images (Table~\ref{HSTposn}). These refined {\it HST} positions were subsequently used to set the astrometric reference  frame for all other images.

We registered the four epochs of Keck imaging to the 2MASS/{\it HST} astrometric grid using between 13-21 stars, solving simultaneously for field center coordinates, rotation, and pixel scale. The solutions yielded offsets relative to the {\it HST} frame of $<$2.5 mas, with a single-axis standard error $\sim$3 mas, and fitting errors of $<0.003^o$ and 0.02\% in field rotation and plate scale, respectively. These results are summarized  in Table~\ref{astrometry}. We averaged the Keck positions of the HST sources for the first three epochs and compared these to the input HST positions, giving $(\Delta\alpha\, ,\Delta\delta)$(Keck-HST)=(-9$\pm5$, 5$\pm8$ mas) with a dispersion of 24 mas predominantly independent of source brightness. We take this dispersion to be representative of the residual uncertainties in the Adaptive Optics PSF across the field, uncertainties in the rotation and plate csale of the HST data (Table~\ref{HSTposn}), residual optical distortion in both the Keck and HST data, and residual refraction in the Keck data. To establish the uncertainty for a particular object we combine this observed dispersion in a root-mean-square sense with the  brightness-dependent term, $\sigma_{SNR}$. This latter term is typically small, $<$10 mas. As noted above,  the 2012 October Keck data were obtained under poor conditions resulting in a dispersion of 50 mas with respect to the HST sources.

For {\it Spitzer} astrometry we used the 4.6 \micron\ (Band 2) Post-BCD mosaics because the source is brighter at this wavelength compared to 3.5 \micron\ and also because of the more symmetric and better sampled PSF at the longer wavelength. We matched {\it HST} positions of $\sim$20 stars (depending on the intersection of the two fields of view), solving  for RA and DEC zero points, rotation angle, and pixel scale. As expected, the {\it Spitzer} scale proved to be highly stable over the four epochs of observation and to differ from the SSC value by $<$0.05\%. We adopted the nominal plate scale for our analysis. We determined the rotation angle to $<0.02^o$ and the positional zero point to $10\sim$20 mas relative to the {\it HST} frame.  We averaged the Spitzer positions of the HST sources for the 4 epochs and compared these to the input HST positions, giving $(\Delta\alpha\, ,\Delta\delta)$(Spitzer-HST)=(7$\pm 9$,-5$\pm 11$) mas with a dispersion of 50 mas predominantly independent of source brightness. We take this value to be representative of the residual uncertainties in PSF  and distortion in both the Spitzer and HST data that must be applied to all sources. We adopt the slightly larger  repeatability measure of $\pm$60 mas obtained in the previous section as the astrometric uncertainty in the Spitzer data; the SNR-related term is negligible by comparison.

\section{Results \label{results}}

\subsection{Determination of Parallax and Proper Motion}

\begin{figure*}
\epsscale{1}\plotone{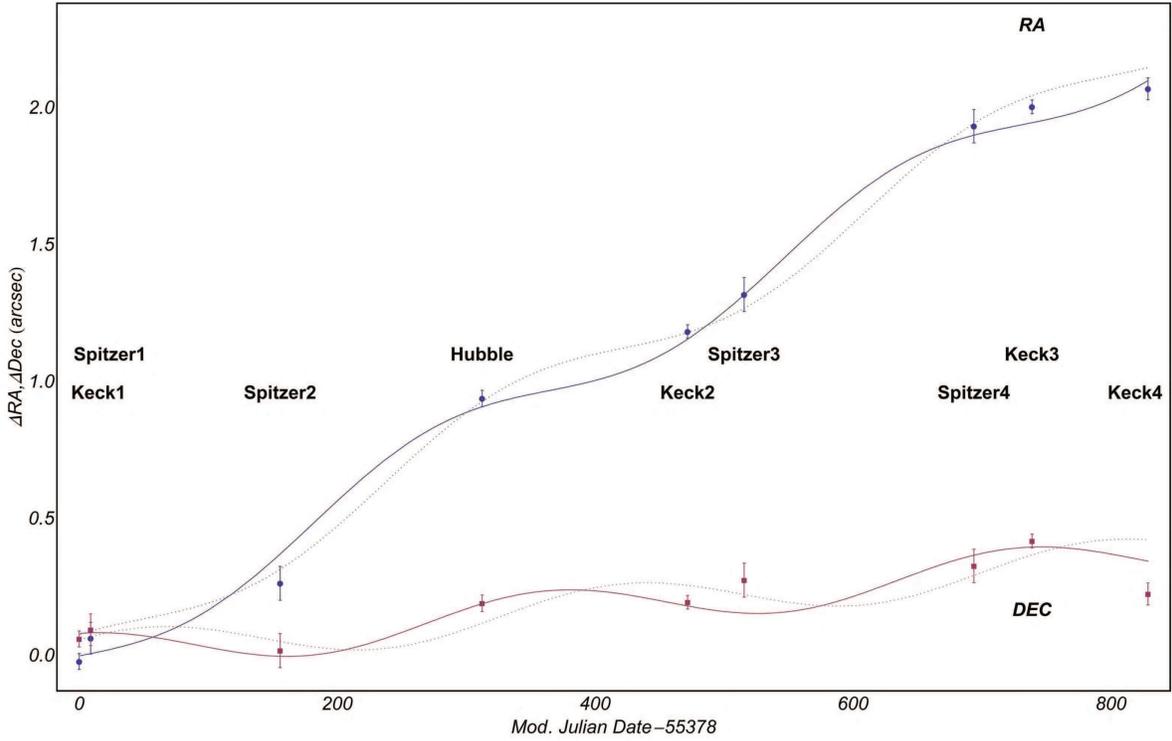} \caption{The fit to the  positions of WISE 1828+2650 obtained using Method-1 over 2 years of observations.  Differences in right ascension ($\Delta\alpha\, cos(\delta$), upper blue lines and circles) and declination ($\Delta\delta$, lower red lines and squares) between WISE 1828+2650 relative to its initial WISE position are shown. Data points are labeled with their appropriate observatory. The solid lines represent the proper motion and parallactic effect as seen from an observatory located on the Earth or in Earth-orbit while the dashed lines represent the view from {\it Spitzer}'s drift-away orbit.\label{AllFit}}
\end{figure*}

\begin{figure*}
\includegraphics[scale=0.5,angle=270]{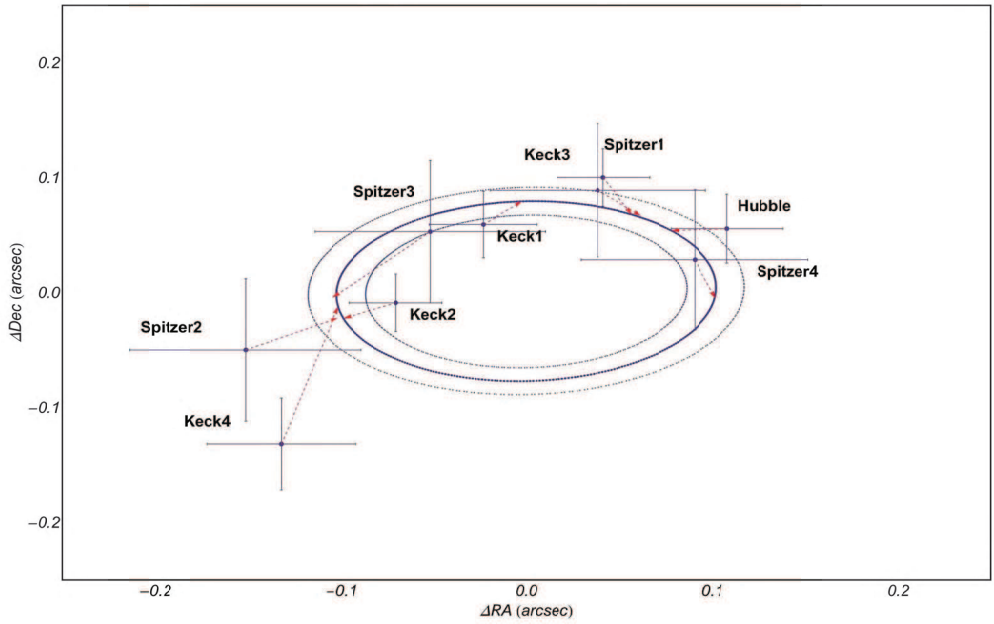} \caption{The fit to the  positions of WISE 1828+2650 obtained using Method-1 over 2 years of observations with the effects of proper motion removed show the fit to the parallactic ellipse. The inner and outer dashed ellipses (blue) show the $\pm$1 $\sigma$ parallax values. Individual sightings are labeled sequentially for each telescope and have an arrow pointing to their expected location on the parallactic ellipse for the time of observation.\label{ellipse}}
\end{figure*}

Table~\ref{astromdata} lists the positions of WISE 1828+2650 on the {\it HST}/2MASS astrometric grid while 
Table~\ref{astromSrcA} lists the offsets of WISE 1828+2650 relative to source A. Both sets of right ascension 
and declination data were fitted to a model incorporating proper motion and parallax \citep{Smart1977, Green1985}:

%
%

\begin{align*}
\alpha^\prime\equiv&\, \alpha_0+\mu_\alpha(t-T_0)/cos(\delta^\prime)\\
\delta^\prime\equiv&\, \delta_0+\mu_\delta(t-T_0)\tag{1}\\
\alpha(t)=&\, \alpha^\prime +\pi\Big( X(t) sin\,\alpha^\prime - Y(t) cos\,\alpha^\prime \Big)/cos\,\delta^\prime \\
\delta(t)=&\, \delta^\prime +\pi\Big( X(t) cos\,\alpha^\prime sin \,\delta^\prime  + Y(t) sin\,\alpha^\prime sin\,\delta^\prime -Z(t) cos\,\delta^\prime \Big)\tag{2} \\
\end{align*}

\noindent where $(\alpha_0, \delta_0)$ are the source position for equinox and epoch $T_0=$J2000.0,
$\mu_{\alpha,\delta}$ are proper motion in the two coordinates in \arcsec/yr, and $\pi$ is the annual parallax. 
The coefficients $X(t),\ Y(t)$, and $Z(t)$ are the rectangular coordinates of a terrestrial or 
earth orbiting telescope (WISE, Keck, {\it HST}) or {\it Spitzer} in its Earth-trailing orbit as seen from the Sun. 
Values of $X,Y,Z$ for the terrestrial or Earth-orbiting observatories are taken from the IDL ASTRO routine 
XYZ. {\it Spitzer} values of $X,Y,Z$ are obtained from the image headers provided by the SSC. Equations (1) and (2) are solved simultaneously using the {\it Mathematica} routine {\it NonLinearModelFit} incorporating appropriate uncertainties for each datapoint. Figure~\ref{AllFit} shows the fit of the data to this model. Figure~\ref{ellipse} shows the fit to parallactic ellipse with the effects of proper motion removed.

We developed two solutions, one using best estimates of the time-varying positions of WISE 1828+2650 in the {\it HST}/2MASS astrometric frame (Method-1; Table~\ref{astromdata}) and the other using offsets between the source and the nearby object denoted Source A (Method-2; Table~\ref{astromSrcA}). The Method-2 position offsets use only the rotation angle and plate scale  derived from the full  astrometric solution, and  should thus be relatively immune to residual distortion and refraction effects that apply across the larger field. On the other hand, the only Method-2 observable is the offset which is compromised by Source-A being both slightly extended in all the images and blended with WISE 1828+2650 in the Spitzer images.

\begin{figure*}
\includegraphics[scale=0.7]{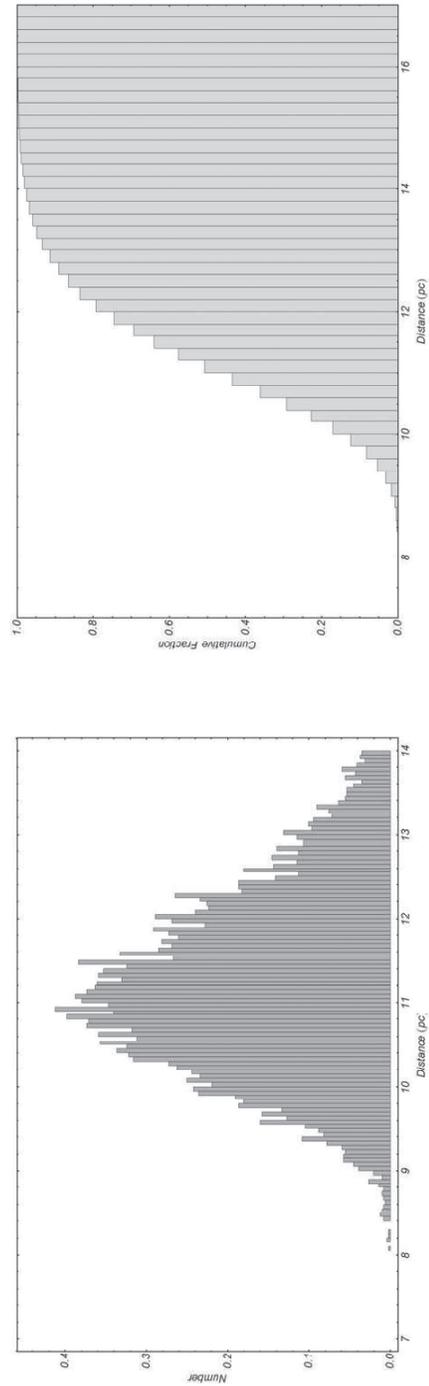} \caption{The histogram (left) represents the results of a 10,000 iteration Monte Carlo calculation for the combined Method-1 and Method-2 datasets wherein the distance solutions were generated from  synthetic data generated with a  Gaussian distribution with appropriate mean and uncertainties (Tables~\ref{astromdata}). The cumulative distribution function on the right gives derived  distances and was used to define the uncertainties in the distance.    \label{histo}}
\end{figure*}

The two solutions give consistent results (Table~\ref{parallax}), yielding parallaxes of 103$\pm$16 mas and 79$\pm$12 mas and corresponding to distances of 9.7$_{-1.1}^{+1.6}$ pc (Method-1) and 12.6$_{-1.9}^{+2.2}$ pc (Method-2). To generate our final kinematic estimates we combined and fitted simultaneously the Method-1 and Method-2 datasets, Method-(1+2). While the input imaging data are the same in both cases, the sources of uncertainty  are different and furthermore,  the blending in the Method-2 offsets lowers the significance of the Spitzer results and thereby halves the number of epochs with precise measurements. Figure~\ref{histo} shows histograms for the parallax and distance  solutions from Method-(1+2) resulting from a 10,000-iteration Monte Carlo calculation wherein the source positions  were varied according to their Gaussian uncertainties and solutions recorded at each step. While the parallax distribution is consistent with the nominal average value and its associated uncertainties, the distance distribution shows a tail to greater distances expected due to its inverse relationship to the parallax. We derived the asymmetric $1\sigma$ distance uncertainties  from the Cumulative Distribution Function of the errors (Figure~\ref{histo}b). Our best estimate of the distance to WISE1828+2650 is 11.2$_{-1.0}^{+1.3}$ pc with a $\chi^2$=36.3 with 31 degrees of freedom. A model of proper motion plus a non-zero parallax provides a much better  fit to the data than a  proper motion-only model which yields  $\chi^2 =$ 135 with 32 degrees of freedom for the Method-(1+2) dataset. On the other hand, there is no evidence that a more complex model, e.g. one including the presence of a perturbing second body in the system, is required. 

Finally, we note that the parallax determination here is a relative one with respect to nearby references stars (not galaxies, except for Source A). Thus the motions of the reference frame can introduce a bias into the parallax determination. Since this bias is at the level of 0.5-2 mas \citep{Dupuy2012}, we can neglect this correction relative to the 100 mas signal of WISE 1828+2650.

\subsection{Spectral Energy Distribution}

\begin{figure*}
\epsscale{0.9}\plotone{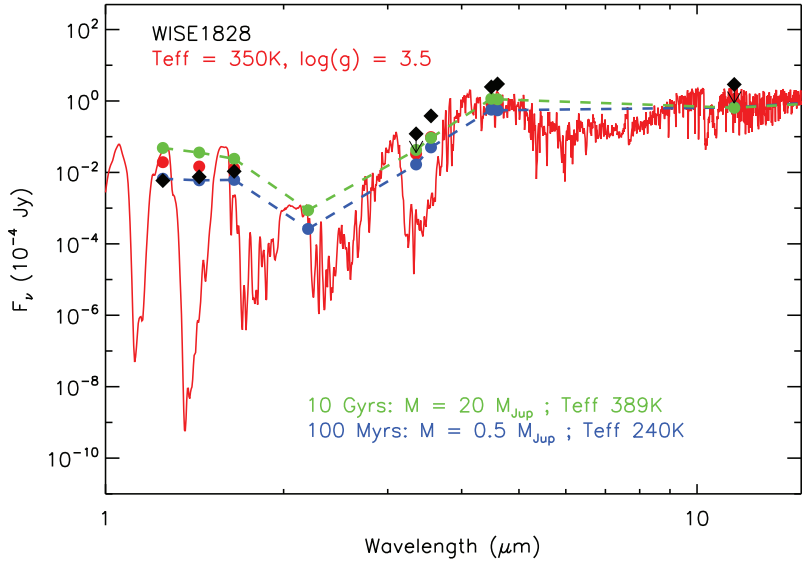} \caption{The spectral energy distribution for the Y dwarf WISE 1828+2650 includes data from {\it HST}, Keck, WISE and {\it Spitzer} (black diamonds). The COND model calculation \citep{Baraffe2003} for a T$_{\rm eff}$ = 350 K object is shown in red, bracketed by the predicted emission in the various bands for (0.1 Gyr,0.5 M$_{Jup}$, 240 K) and (10 Gyr,20 M$_{Jup}$, 389 K) objects with green and blue circles, respectively. 
\label{SED}}
\end{figure*}

WISE 1828+2650 is one of the coolest (latest) Y dwarfs yet identified. Available photometry of WISE 1828+2650 covers the 1-5 $\mu$m range and can compared against the predictions of brown dwarf atmospheric and evolutionary models \citep{Baraffe2003}. Figure~\ref{SED} shows a number of models fitted to the observations as described below. Unfortunately, none of these models provide a good fit to the full range of the data. The COND models pass through either the two short wavelength points or the 3-5 \micron\ points, but not both. The over- or under-estimates are significant, up to factors of 5, which suggests that the models have omitted one or more important physical processes in the atmosphere, e.g. non-equilibrium chemistry or new absorbers, that dramatically affect the SED. \citet{Marley2010} highlight the importance of clouds in the L/T dwarf transition which greatly change the near-IR (J, H and K band) appearance of these warmer objects. The onset of still lower temperature condensates, e.g. Na$_2$S producing sulfide clouds    \citep{Morley2012}, may play a role in the appearance of the Y dwarfs (see however \citet{Burrows2003}).

While acknowledging the failings of the models in this uncharted range of effective temperature, we will use them  to make preliminary estimates of the properties of WISE 1828+2650. We  take advantage of our knowledge of the distance to the source and use absolute magnitudes rather than relative fluxes to constrain further the evolutionary status of the object.

\begin{figure*}
\epsscale{0.8}\plotone{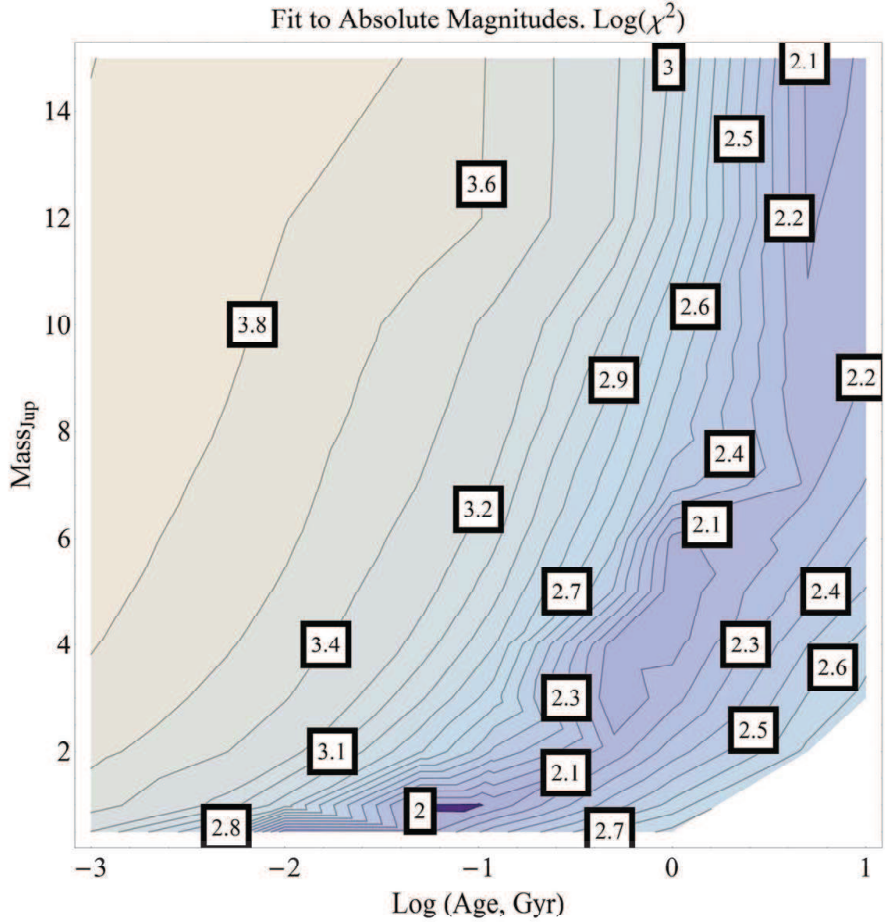} \caption{The quality of fit ($\chi^2$)  between the photometric observations (Keck H-band, HST F140W,  Spitzer channels 1 and 2)  and the COND models is shown as a function of Age (Gyr) and Mass (M$_{Jup}$). The contours give the logarithm of the $\chi^2$ of the fit for 2 degrees of freedom (4 data points minus  two derived values). As discussed in the text, the best fit is for a very young, low mass object which is not likely on dynamical grounds. For an adopted dynamical age of 2-4 Gyr, the best fitting mass is in the range of 3-6 (M$_{Jup})$.
\label{BDchi2}}
\end{figure*}

Magnitudes corresponding to the dust-free {\it COND} models \citep{Baraffe2003} were calculated using bandpasses matched to {\it HST}, WISE, {\it Spitzer}, and ground-based filters. Models were first selected to match either the absolute H magnitude M$_H=22.21^{+0.25}_{-0.22}$  mag  or M$_{[4.5]}$=14.42$^{+0.24}_{-0.20}$ mag where the uncertainty in the absolute magnitude is dominated by the  distance uncertainty.  Selecting models matching either absolute magnitude led to two classes of object. Matching the H band absolute magnitude leads to  atmospheres with a narrow range of effective temperatures (T$_{\rm eff} = 275\pm 40 $K) but with a broad variation of ages and masses from (0.5 M$_{Jup}$, 0.05 Gyr) to (8 M$_{Jup}$, 10 Gyr). Matching the [4.5$\mu$m] absolute magnitude leads to hotter, more massive objects (T$_{\rm eff}=450\pm 40 $K) with a range of ages and masses from (2 M$_{Jup}$, 0.1 Gyr)  to (30 M$_{Jup}$, 10 Gyr).  Matching the absolute magnitudes in F140W, H, and the two Spitzer bands {\it simultaneously}, each with an uncertainty of $\sim$0.25 mag, yields a best fitting model with  (0.1 M$_{Jup}$, 0.1 Gyr),  T$_{\rm eff}=285$K, and a $\chi^2$=95 for 2 degrees of freedom (4 data points and two parameters). The variation of the quality of fit as a function of mass and age (Figure~\ref{BDchi2}) shows a minimum for a young, low  mass object, but a broad valley tending to older, more massive objects.  We emphasize the poor quality of the model fits by pointing out that the contour values in Figure~\ref{BDchi2} show the logarithm of the $\chi^2$ value.

Finally, we examined the possibility that the SED could be the result of a combination of two coeval bodies in a binary system \citep{Leggett2012}. Fitting the 4 absolute magnitudes yields a pair of equal mass objects with (0.5 M$_{Jup}$, 0.05 Gyr),  T$_{\rm eff}=285$K, and $\chi^2$ = 55. However, since none of these models provides a particularly good fit to the data across the entire 1-5 \micron\ range, the estimated physical parameters should be taken with a grain of salt. What all the solutions have in common is a low T$_{\rm eff}\sim250-450$K and a mass range of a 0.5-20 M$_{Jup}$ depending on the unknown age of the system ($\S$\ref{ages}). 

Other models lead to a similar range of predicted properties for WISE 1828+2650 depending on the wavelength used for the comparison and the assumed mass and/or age. The \citet{Marley2010} and \citet{Burrows2004} models also suggest an effective temperature $\sim$ 300-400 K which can be associated with a broad range of masses and ages with no single model providing a good simultaneous fit to all of the 1-5 $\mu$m observations. \citet{Spiegel2012} calculate models for both ``hot start'' and ``cold start" formation scenarios \citep{Marley2007} for ages less than 0.1 Gyr. Matching our data to the predicted absolute $H$ band models for either the ``cold start'' or ``hot start'' cases yields 1 M$_{Jup}$ objects with ages of 0.02-0.03 Gyr. Matching  the predicted  $M$ band models (only an approximation to our Spitzer filter) yields objects in the range (1-5 M$_{Jup}$, 0.007-0.1 Gyr). Differences between the COND models and the Spiegel \& Burrows models are relatively minor in the area of age overlap.

\section{Discussion}

\subsection{Physical Properties of WISE 1828+2650 \label{physical}}

The nature of WISE 1828+2650 is ambiguous for two reasons. First, the failure of the models to reproduce successfully the SED and absolute magnitudes makes it hard to draw any quantitative conclusions. Second, even with  improved models, e.g. ones including clouds or new low-temperature absorbers,  there would remain a family of models with roughly similar T$_{\rm eff}$ that are degenerate in mass and age.

WISE 1828+2650 is similar in appearance to the brown dwarf companion to the $\sim2$ Gyr old white dwarf WD 0806-661 \citep{Luhman2012} located 19.2 pc away. The twofold greater distance to this system has so far prevented a detection of WD 0806-661 B in the 1-2 $\mu$m region. As a result, we do not yet have a comparable SED, but the near- to mid-IR colors are comparably red with J-[4.5]=9.25$\pm$0.35 \citep{Kirkpatrick2012} for the WISE object and J-[4.5]$>$7 for WD 0806-661 B. The 3-5 $\mu$m colors are also similar, [3.6]-[4.5]= 2.77$\pm0.15$ mag for WD 0806-661 B and [3.6]-[4.5]= 2.81$\pm0.06$ mag for WISE 1828+2650, although neither of these objects are as red as other Y dwarfs. With an age estimate for the white dwarf it is possible to narrow the mass range for its companion based on its colors. \citet{Luhman2012} suggest a mass of 6-9 M$_{Jup}$ which is consistent with our estimates for WISE 1828+2650 in that age range.

\begin{figure*}
\plottwo{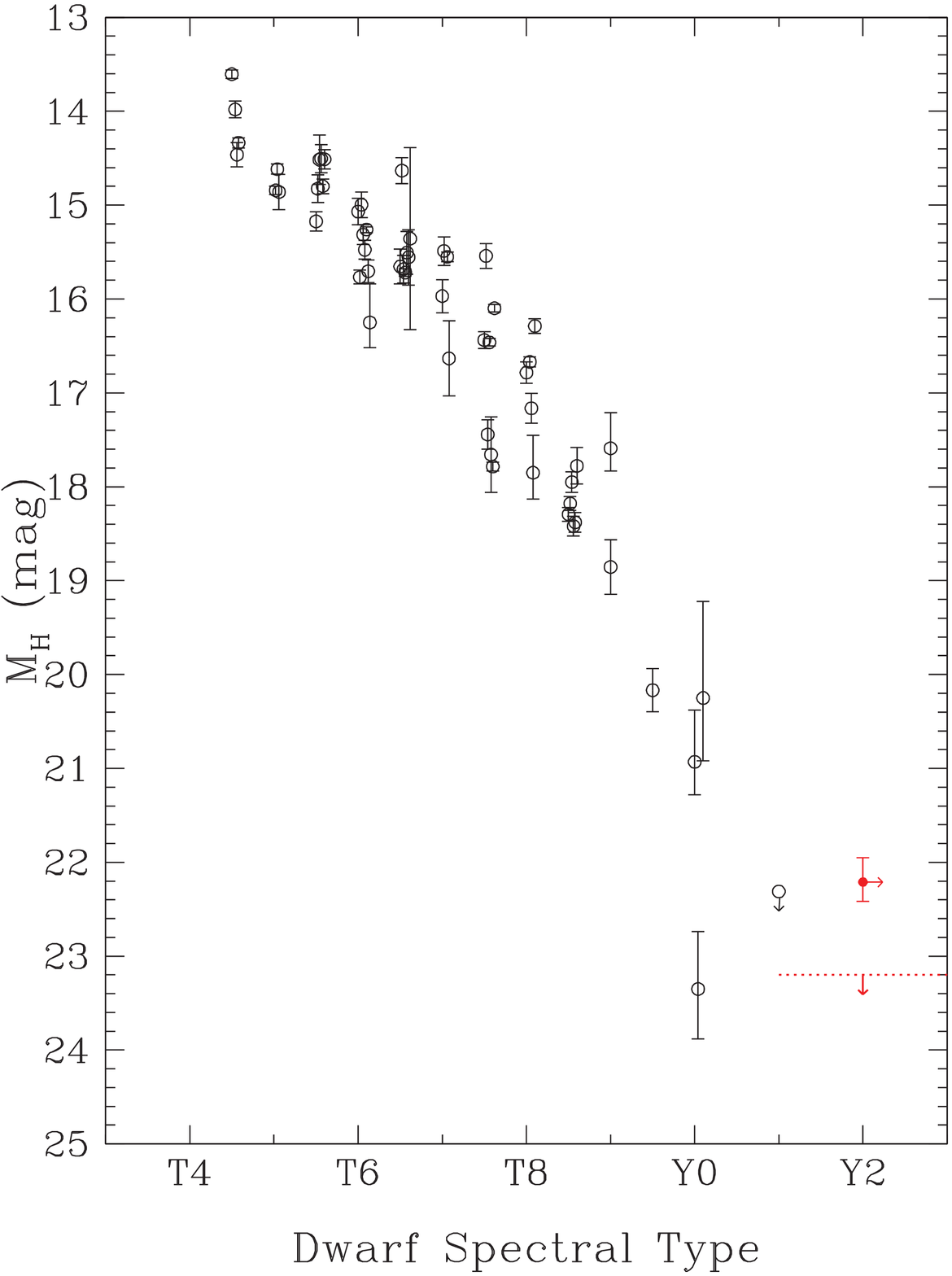}{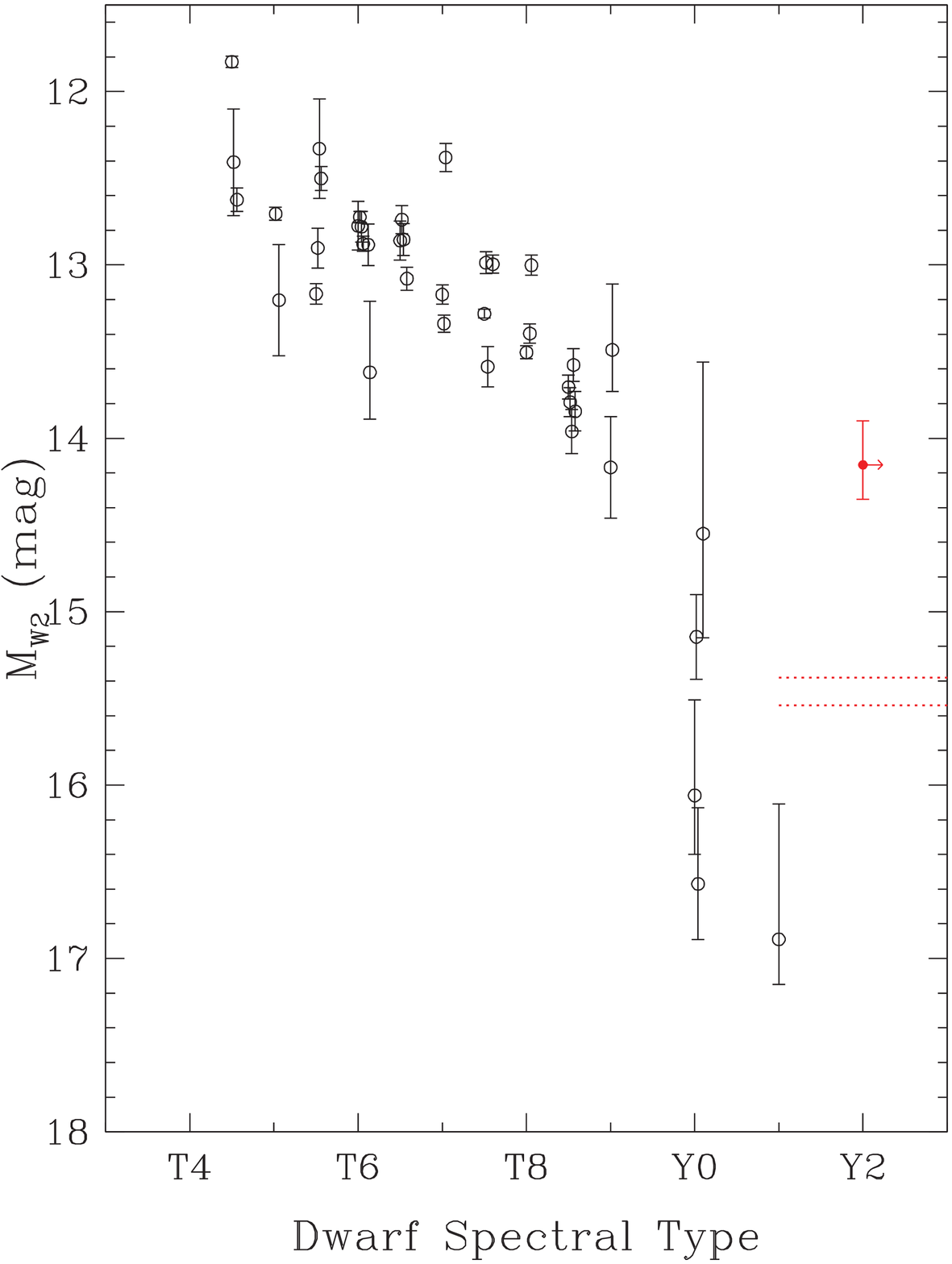} 
\caption{Absolute magnitude versus spectral type diagrams for late-T and Y dwarfs at the H-band (left) and WISE W2-band (right) for objects with 
secure parallaxes. The figures are adapted from \citet{Kirkpatrick2012} and include revised parallax measurements as reported
in \citet{Marsh2012}. A new Y dwarf (WISE J163940.83-684738.6) from \citet{Tinney2012}
 has  been added to the M$_{W2}$ vs type plot. Shown by the red, solid point is the location of WISE 1828+2650. Dashed red lines mark the absolute magnitude upper limit at H-band and the one-sigma upper and lower limits in the W2-band for WD 0806-661B. Because there is no spectrum for WD 0806-661B, the spectral type range shown is an estimate. (For this source, the H magnitude limit is based on the J magnitude limit given by \citet{Luhman2012}, assuming the J-H color of 0.7 from WISE 1828+2650; WISE does not detect this source, so the IRAC channel 2 magnitude, which is nearly equivalent to the W2 magnitude, is plotted instead.) Note that at W2 in particular, both WISE 1828+2650 and WD 0806-661B lie significantly above the trend established for earlier Y dwarfs.
\label{CMD}}
\end{figure*}

An important peculiarity  of WISE 1828+2650 is its bright absolute magnitude, M$_{W2}$ = 14.15$^{+0.25}_{-0.21}$ mag for its presumed Y2 spectral type. Compared with WD 0806-661 B  with an absolute magnitude of M$_{[4.5]}$=15.5 mag, WISE 1828+2650 is $>$ 1 mag more luminous.  More troubling, however, is the fact that compared with other Y dwarfs with reasonably secure parallaxes, WISE 1828+2650 is $>$ 2 mag brighter than other Y dwarfs in M$_{W2}$ and, to a lesser extent, brighter than some of the earlier Y dwarfs in M$_H$. Figure~\ref{CMD}, which illustrates this problem, is an updated version of the absolute magnitude vs. spectral type figures appearing in \citet{Kirkpatrick2012}. This new figure uses recently published Y dwarf parallaxes by \citet{Marsh2012} and \citet{Tinney2012}. The locations of WISE 1828+2650 and WD 0806-661 B are also shown. The spectral type of WD 0806-661 B has been estimated from its effective temperature since no spectrum is available and only an upper limit is available for the H-band magnitude. However, in both cases it is apparent that the two objects are similarly deviant from the trend set by earlier Y objects.

A simple explanation of at least part of the $\sim$1.5 mag difference in absolute  M$_{[4.5]}$ between WISE 1828+2650 and WD 0806-661 B is that WISE 1828+2650  is an unresolved, equal mass binary which would increase the total system brightness by 0.7 mag, close to the difference with WD 0806-661B.  Any such companion to WISE 1828+2650 would have to have an orbit $<$ 0.5 AU ($<$0.05\arcsec) to be consistent with the HST and Keck data and if present would eventually manifest itself in either the astrometric data or future spectroscopic data. No such evidence for binarity yet exists. In any event, such an explanation does not account for the significant deviation from the trends established for earlier spectral types. Some other explanation must be sought for the differences in absolute magnitude at 3-4 \micron. 

\subsection{Age of WISE 1828+2650\label{ages}}

We cannot eliminate solutions corresponding to young, low mass objects for WISE 1828+2650. Of the approximately 250 low mass objects (M, L and T spectral types) within 10 of the Sun there are at least three M stars younger than $\sim$100 Myr: AU Mic (9.9 pc; \citet{Kalas2004}), GJ 896 B (6.3 pc; \citet{Riedel2011}), and most recently AP Col (4.6 pc) \citep{Henry2006, Riedel2011}. A number of  brown dwarfs  have been claimed to be young based on near-IR spectra which indicate low surface gravity. Examples include the L dwarf 2MASS J01415823-4633574 (Kirkpatrick et al. 2010; Patience et al. 2012) and  the nearby, 7.5 pc, L5 brown dwarf 2MASS J035523+113337   \citep{Faherty2012}. If these 2MASS objects are  indeed that young, then their  masses might be as low as 10 M$_{Jup}$. As noted above, if WISE 1828+2650 were as young as 0.1 Gyr, its mass would be only 1-2 M$_{Jup}$.  A young, brown dwarf of this age is certainly  a possibility, albeit one with a low probability ($\sim$1\%) based on the statistics of nearby low mass objects, i.e. 3 young M stars plus 1 young L dwarf out of 250 low mass objects.

The simplest metric for assessing the age of WISE 1829+2650 is the directly measured tangential velocity which comes from the proper motion and distance: $v_{tan}=4.74 \mu/\Pi=51 \pm 5$ km s$^{-1}$ where $\mu$ is the total proper motion and $\Pi$ the parallax  \citep{Smart1977}. This value falls in the middle of the distribution of tangential velocities measured for nearby  ($<$ 20 pc) L and T brown dwarfs (see Figure 7, \citet{Faherty2009}). These velocities are consistent with models of disk population with ages of 3-8 Gyr. This range  narrows to  2-4 Gyr if the highest velocity objects ($v_{tan}>100$ km s$^{-1}$ are excluded.  Within the age range and using fits to the COND models described above, we estimate that the mass of WISE 1828+2650 is $\sim$ 3-6 M$_{Jup}$.

A number of authors have used kinematic properties  to associate  brown dwarfs with nearby young clusters, but  the results to date are only suggestive   \citep{Faherty2012}. For example, might WISE 1828+2650 be a young object could be associated with one of the nearby young moving groups, e.g. AB Dor, $\beta$ Pic or Tucana/Horlorgium   \citep{Zuckerman2004,Torres2008} or the Argus/IC 2391 group as in the case of AP Col? These groups are located at distances of 20-100 pc with ages ranging from 8 to 50 MYr. We evaluated possible galactic velocity vectors for WISE 1828+2650, $(U,V,W)$, and their associated uncertainties $\sigma_{U,V,W}$, from its position, parallax, and proper motion vector determined here for a range of the {\it unknown} $V_{rad}$ from -100 to 100 km s$^{-1}$. We calculated the difference between cluster $(U_{cl},V_{cl},W_{cl})$ values and possible values for WISE 1828 as a function of $V_{rad}$: 

\begin{align*}
\Delta UVW(V_{rad}) 	&  = \sqrt{(U_{WISE}(V_{rad})-U_{cl})^2+(V_{WISE}(V_{rad})-V_{cl})^2  +(V_{WISE}(V_{rad})-V_{cl})^2} \tag{3} \\
\end{align*}

For V$_{rad}$ between -100 to 100 km s$^{-1}$ there is no good match in $(U,V,W)$ space for any of the six  moving groups with ages from 8 to 50 Myr (Table~\ref{UVW};  \citet{Zuckerman2004b};  \citet{Riedel2011}). The smallest separation is $\Delta UVW\sim50$ km s$^{-1}$. While it is impossible to rule out the possibility that WISE 1828+2650 is a young  straggler ejected from one of these groups, it is more likely that WISE 1828+2650 is an older, higher mass object.

\subsection{Brown Dwarf or Free Floating Planet?}

The suggestion that a population of free floating planets may be responsible for short duration microlensing events, $t\sim \sqrt{M/M_{Jup}}$ day $\sim 0.3-3$ day for M=0.1-10 M$_{Jup}$ \citep{Sumi2011} immediately leads to the conjecture that objects like WISE 1828+2650  might be members of such a population. \citet{Sumi2011} calculate that fitting the distribution of short duration microlensing events requires a population of $\sim$1-2 M$_{Jup}$ objects roughly 1.3-1.9$\times$ as numerous as all main sequence stars (0.08-1 M$_\odot$). \citet{Quanz2012} argue that imaging surveys may be starting to reveal a significant number of bound planets located at 10s of AU from low mass stars which would decrease the need for a large population of free-floating, planetary mass objects.  

The RECONS database \citep{Henry2006} lists 300 main sequence stars (M-G) within 10 pc, or 0.072 object pc$^{-3}$. Estimates of the entire population of Y dwarfs down to the coolest objects are premature but we note that WISE 1828+2650's distance implies a minimum local density of   $1.7\pm0.5\times10^{-4}$ pc$^{-3}$. For WISE 1828+2650  to represent the tip of the microlensing iceberg, the population of similar objects would have to rise very sharply below the few $M_{Jup}$ mass of this object. Between 0.1 and 10 M$_{Jup}$ where the \citet{Sumi2011} data are sensitive (Figure 9 of \citet{Sumi2011}), the mass distribution function, $dN/DM\propto M^{-\alpha}$, would have to have to be quite steep. Estimating  the mass of WISE1828+2650 using a uniform distribution between 3-6 $M_{Jup}$ yields  $\alpha=2.2\pm 0.2$ to produce enough extremely low mass  dwarfs (0.1-10 M$_{Jup}$) equal in number to the main sequence stars within 10 pc.  This is dramatically steeper than the flat Brown Dwarf index, $\alpha\sim0$ (Figure 14 in \citet{Kirkpatrick2012}), and steeper than the mass distribution of planets determined by radial velocity studies, $\alpha=1.5$ \citep{Cumming2008, Howard2010}.

To date, fourteen ``free-floating" Y dwarfs have been identified \citep{Cushing2011, Kirkpatrick2012, Tinney2012}. Several other WISE-selected Y dwarf candidates, such as those presented in \citet{Griffith2012}, remain but lack spectroscopic confirmation. All are on active parallax and spectroscopic classification campaigns, so it will soon be known if other WISE 1828+2650 and WD 806-661 B analogs exist. Even if, for example, the local density of objects as cool as WISE 1828+2650 eventually proves to be 5 times higher than estimated solely on the basis of this one object, then the power law index required to account for the microlensing results would still be $>$ 1.5. In the absence of convincing evidence for an upturn in the mass distribution, the most conservative interpretation is that WISE 1828+2650 sits at the low mass end of the brown dwarf distribution rather than at the start of a much larger population of free floating ``planets".

A less conservative interpretation of these results is also possible. As Luhman et al. (2011) themselves suggest WD 0806-661 B could have formed as a $\sim$6-9 M$_{Jup}$ object in a protoplanetary disk around its $\sim$2 M$_\odot$ primary, WD 0806-661 A. Only later, after the primary shed mass during its evolution off the main sequence, would WD 0806-661 B have migrated into a larger orbit, currently of radius $\sim$2500 AU. The formation mechanism between the WISE Y0 and Y1 discoveries, which presumably formed via cloud fragmentation as do normal stars, and WB 0806-661 B would thus be different. Compositional differences should be seen between the two types of objects, given that chemical differentiation is expected in a proto-planetary disk and not in a star formation cloud. Jupiter's enhanced metallicity is an obvious example of this difference \citep{Wong2004}. Objects formed via core accretion may also have a pronounced core of heavy elements as inferred for transiting planets \citep{Rogers2010} which  might  manifest itself in significant ways. Such differences between these  two classes could possibly account for the discrepancies noted in the absolute magnitude versus spectral type diagrams of Figure~\ref{CMD}.

Taking this one step further, the similarity between WD 0806-661 B and WISE 1828+2650 may be a direct reflection of the fact that both objects were birthed via the same process. In this case, WISE 1828+2650 would be a  ``free-floating" planet that, given our mass estimate of $\sim$3-6 M$_{Jup}$, would fall at the high-mass end of the rogue planet population proposed by \citet{Sumi2011}. In this case, the poor agreement with theoretical spectra may be a consequence of a poor match between the true elemental abundances or interior structure of WISE 1828+2650 and those in the model assumptions. If the space density of truly ``free-floating" planets measured by  \citet{Sumi2011} is correct, then there should be colder examples ($<$300K) at much smaller distances ($<<$10 pc) than WISE 1828+2650 itself. Some of these may yet be  discoverable at lower signal-to-noise ratios in the WISE data.


\section{Conclusions}

We have used images from four telescopes to determine the parallax and distance to the $>$Y2 dwarf WISE 1828+2650. At $11.2^{+1.3}_{-1.0}$ pc, the object is among the coolest, lowest luminosity objects yet discovered in the solar neighborhood. The high tangential motion of the object suggests an age of 2-4 Gyr with a mass of 3-6 M$_{Jup}$. However, the absolute magnitude  of WISE 1828+2650 is brighter by several magnitudes (depending on wavelength) than extrapolated from other Y dwarfs making the true nature of this source somewhat of a mystery. WISE 1828+2650 is similar  in color and absolute magnitude to the cool object orbiting the nearby white dwarf WD 0806-661 B. The exact nature and evolutionary state, including its  mass and age, of WISE 1828+2650 will require further observation and theoretical investigation.  

The proximity of WISE 1828+2650 and other Y dwarfs will allow  more precise determinations of their  motions in the coming years as well as a search for any astrometric companions that might fall below detectability via direct imaging. The launch of  JWST will allow measurement of the spectrum of WISE 1828+2650  and other objects with similarly low T$_{\rm eff}$ from the deep red through to the mid-infrared,  resulting in a much more detailed physical characterization and  a better assessment of their ages and masses. Y dwarfs like WISE 1828+2650 and the companion WD 0806-661 B will become touchstones for spectroscopic modeling with direct applicability to the atmospheres of bona-fide planets now being measured with transit spectroscopy and direct imaging, e.g. \citet{Barman2011}. The differences expected between bound objects (``planets") and free-floating objects (likely brown dwarfs, but possibly free floating ``planets") should produce distinct spectroscopic markers distinguishable in the high SNR, high spectral resolution measurements possible with JWST.

\acknowledgments

The research described in this publication was carried out in part at the Jet Propulsion Laboratory, California 
Institute of Technology, under a contract with the National Aeronautics and Space Administration. 
This publication makes use of data products from the Wide-field Infrared Survey Explorer, which is a 
joint project of the University of California, Los Angeles, and the Jet Propulsion Laboratory/California 
Institute of Technology, funded by the National Aeronautics and Space Administration.
 This research has made use of the NASA/IPAC Infrared Science Archive (IRSA),
which is operated by the Jet Propulsion Laboratory, California Institute of Technology, under contract
with the National Aeronautics and Space Administration.
This work is based in part on observations made with the {\it Spitzer} Space Telescope, which is
operated by the Jet Propulsion Laboratory, California Institute of Technology, under a contract with
NASA. Support for this work was provided by NASA through an award issued to program 70062 and 80109 by JPL/Caltech. This work
is also based in part on observations made with the NASA/ESA {\it Hubble} Space Telescope, obtained
at the Space Telescope Science Institute, which is operated by the Association of Universities for
Research in Astronomy, Inc., under NASA contract NAS 5-26555. These observations are associated with 
program 12330. Support for program \#12330 was provided by NASA through a grant from the Space
Telescope Science Institute. Some data presented herein were obtained at the W. M. Keck Observatory from telescope time allocated to the National Aeronautics and Space Administration through the agency's scientific partnership with the California Institute of Technology and the University of California. The Observatory was made possible by the generous financial support of the W. M. Keck Foundation. The authors wish to recognize and acknowledge the very significant cultural role and reverence that the summit of Mauna Kea has always had within the indigenous Hawaiian community. We are most fortunate to have the opportunity to conduct 
observations from this mountain. In addition we wish to acknowledge the generosity of Jay Anderson in sending {\it HST} 
data for the clusters M15 and M92 and useful discussions with Andy Gould. The RECONS database of nearby stars remains
an invaluable resource. We used the IRSA archive at IPAC to access the 2MASS catalogs.

\section{Appendix I. Differential Refraction and Distortion Mapping}

\begin{figure*}\includegraphics[scale=0.15,angle=90]{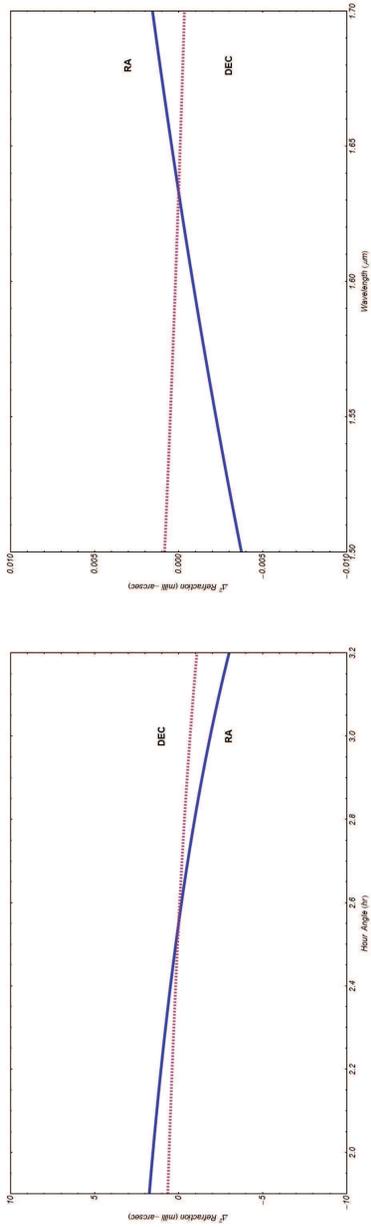}\caption{ Effects of 2$^{nd}$ order differential refraction as a function of hour angle and central wavelength for observations of 1828+2650. \label{DiffRefrac}}
\end{figure*}

The positions of sources in the Keck images were corrected for the effects of differential refraction. The error introduced by coadding before source extraction is small compared to other sources of astrometric error. We applied to each extracted position a correction $\Delta_{\alpha,\delta}=\Delta_{\alpha,\delta}(HA_0,\delta_0,\lambda_0)- \Delta_{\alpha,\delta}(HA+\Delta HA,\delta_0+\Delta\delta,\lambda_0)$ relative to the center of the field where $HA_0$ and $\delta_0$ are the hour angle and declination at the center of the field with the HA specified at the middle of the roughly hour-long observational sequence.  $\lambda_0$ is the central wavelength of the H-band filter. Figure~\ref{DiffRefrac} shows 2$^{nd}$ order corrections to the  differential refraction as a function of Hour Angle during the duration of one observation and for a change in effective wavelength across the H band. The curves show the changes in the  differential refraction from the nominal 1$^{st}$ order correction at one corner of the Keck field of view ($\sim$20$\sqrt{2}$\arcsec, center to corner). At the level of precision required for these measurements, these 2$^{nd}$ order differential effects can safely be ignored over the small Keck field. 

\begin{figure*}\includegraphics[scale=0.5]{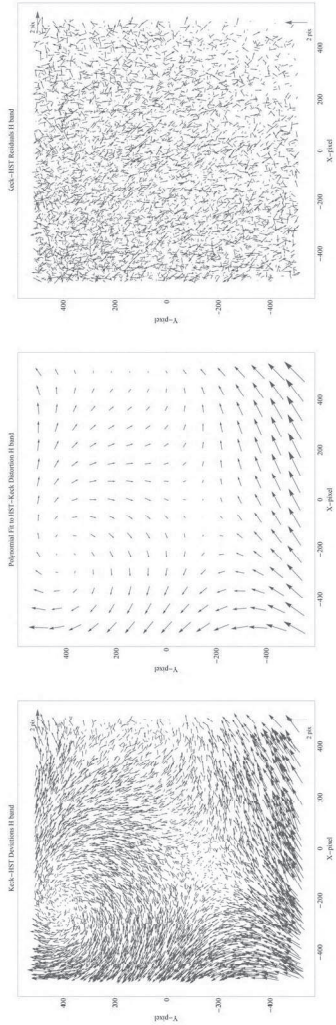}\caption{NIRC2-LGS observations of the cluster M15 were compared with observations taken with HST/WFC3 \citep{Anderson2008}. left) The difference between the HST and Keck position as a function of position in the Keck focal plane; center) a 4${th}$ order polynomial fit to the distortion map; right) residuals remaining after subtracting the model from the data.\label{distortion}}
\end{figure*}

To assess the effects of optical distortion on NIRC2 images in the wide field mode (40 mas pixels) we followed the procedure outlined in \citet{Yelda2010} using {\it HST} observations of the cluster M15. The HST data \citep{Anderson2008} were corrected  for the effects of non-orthogonal axes (coordinate skew). Keck observations of M15 were made in the H band using NIRC2 with a natural guide star Adaptive Optics at 2 epochs (Oct 2011 and July 2012). In July 2012 observations were made at 3 different orientations of the NIRC2 camera (0$^o$,$\pm$45$^o$). At each rotation angle and epoch the telescope was stepped through a 9 position pattern across the cluster for a total of 36 separate observations of some portion of the cluster with each image containing between 50-250 matches between the Keck and HST source lists. The Keck data were extracted using IDL aper routine with a 5 pixel radius. After correction for the effects of differential refraction, lists of sources extracted from each image were rotated and shifted into registration with the HST stars. In total the HST stars probed 5631 unique $(x,y)$ positions in the NIRC2 focal plane (Figure~\ref{distortion}). Nominal $(\alpha,\delta)$ positions were obtained for the Keck sources using plate scale and rotational information derived from the image registration. Noise-weighted differences between the Keck and HST positions, $(\Delta\alpha,\Delta\delta)$,  were fitted to a fourth order polynomial as a function of focal plane coordinates $(x,y)$ using the {\it Mathematica} routine {\it LinearModelFit}. The derived coefficients for transformation from native, distorted $(x,y)$ pixel coordinates into un-distorted, $(x^\prime,y^\prime)$ pixel coordinates are given in Eqns.~4-5 and Table~\ref{XYmap}. The coordinate system is centered at the center of the NIRC2 array and extends from $\pm$512.

\begin{align}
x^\prime=&a_0 + a_1 x+ a_2 y + a_3 x^2+a_4x y +a_5 y^2+ a_6 x^3 +a_7x^2 y \nonumber\\
&a_8 x y^2+ a_9  y^3+ a_10 x^4+ a_11 x^3 y+ a_12 x^2 y^2+ a_13 x y^3+y^4 \tag{4}\\
y^\prime=&b_0 + b_1 x+ b_2 y + b_3 x^2+b_4x y +b_5 y^2+ b_6 x^3 +b_7x^2 y\nonumber\\
&b_8 x y^2+ b_9  y^3+ b_10 x^4+ b_11 x^3 y+ b_12 x^2 y^2+ b_13 x y^3+y^4 \tag{5}
\end{align}

Standard equations then project the  $(x^\prime,y^\prime)$  pixel coordinates onto the sky accounting for plate scale  and rotation angle. Our H-band plate scale of 39.765$\pm$0.004 mas/pixel is consistent with the 39.82$\pm$0.25 mas/pix from \citet{Metchev2004} from comparison with a known binary system and somewhat smaller than the 39.90$\pm$0.08 and 39.86$\pm$0.02 K$_s$plate scales derived separately in x- and y- by  \citet{Pravdo2006} from comparison with 17 HST stars.

Examination of the residuals led to the rejection of $<$2\% of the data, i.e. 100 points that deviated significantly in either amplitude or orientation from their neighbors. The first plot of the Figure~\ref{distortion} triptych shows Keck-HST positions differences as vectors. The central image shows a vector representation of the 4$^{th}$ order polynomial fitted to the distortion map. The third image shows the residuals in the Keck-HST differences after subtraction of the polynomial fit. The mean value of the Keck-HST residuals after the model subtraction in focal plane coordinates is $(x,y)=(0.2\pm10,1.0\pm13)$ mas. The distortion map was applied to the individual Keck images using an IDL script for ``de-warping'' developed by the Keck Observatory\footnote{http://www2.keck.hawaii.edu/inst/nirc2/forReDoc/post\_observing/dewarp/}.

\clearpage

\begin{deluxetable}{lccr}
\centering
\tablecaption{ Observing Log \label{ObsLog}}
\tablehead{
\colhead{Telescope}&\colhead{Epoch}&\colhead{Modified Julian Date}&\colhead{Comment}}
\startdata
WISE 			& 2010-03-30 	&55285.00&\cr
Keck/NIRC2-LGS 		& 2010-07-01 	&55378.44&PI, Wright\cr
{\it Spitzer}/IRAC	& 2010-07-10 	&55387.34&PI, Mainzer, PropID=9/551\cr
WISE 			& 2010-09-28 	&55467.00&\cr
{\it Spitzer}/IRAC	& 2010-12-04 	&55534.27&PI, Kirkpatrick, PropID=10/70062\cr
{\it HST}/WFC3-IR 	& 2011-05-09 	&55690.89&PI, Kirkpatrick,PropID=12330\cr
Keck/NIRC2-LGS		& 2011-10-16 	&55850.21&PI, Beichman\cr
{\it Spitzer}/IRAC	& 2011-11-19	&55894.04&PI, Kirkpatrick,PropID=10/80109\cr
{\it Spitzer}/IRAC	& 2012-05-25	&56072.25&PI, Kirkpatrick,PropID=10/80109\cr
Keck/NIRC2-LGS		& 2012-07-09 	&56117.32&PI, Beichman\cr
Keck/NIRC2-LGS		& 2012-10-07 	&56207.22&PI, Beichman\cr
\enddata
\end{deluxetable}

\begin{deluxetable}{lcccccr}
\tiny
\centering
\tablecaption{HST Sources in Other Datasets  \label{HSTposn}}
\tablehead{
\colhead{RA}&\colhead{DEC}&\colhead{F140W}&\colhead{$\sigma$}&\colhead{Center}&\colhead{Posn.}&\colhead{Dataset}\\
\colhead{(J2000)}&\colhead{(J2000)}&\colhead{ (mag)}&\colhead{(mag)}&\colhead{Dist.(\arcsec)}&\colhead{Err. (mas)}&}
\startdata
277.10529566&26.85705074&15.66&0.001&84.9&42.3&Spitzer,2MASS\\
277.10824423&26.85082312&19.56&0.020&65.7&32.8&Spitzer\\
277.11519519&26.83369321&17.27&0.002&52.5&26.2&Spitzer\\
277.11647015&26.84300085&16.57&0.002&34.2&17.0&Spitzer,2MASS\\
277.11790606&26.85103552&19.63&0.009&39.6&19.7&Spitzer\\
277.12026174&26.82339456&17.37&0.002&76.3&38.0&Spitzer\\
277.12155151&26.84341695&18.21&0.003&17.8&8.9&Keck\\
277.12463494&26.84734988&19.24&0.005&15.3&7.7&Keck\\
277.12479343&26.84571497&21.22&0.018&10.3&5.6&Keck\\
277.12494638&26.84401162&20.30&0.010&7.0&3.7&Keck\\
277.12662489&26.84369877&21.76&0.025&1.5&3.3&Keck\\
277.12761710&26.86279933&17.29&0.002&68.8&34.2&Spitzer\\
277.12799324&26.84048725&19.13&0.005&11.9&6.0&Keck\\
277.12816179&26.84302297&20.64&0.013&4.2&2.7&Spitzer,Keck\\
277.12865656&26.83903755&18.24&0.003&17.5&8.7&Keck\\
277.12877327&26.85757557&15.28&0.001&50.2&25.0&Spitzer,2MASS\\
277.12906930&26.84100634&19.52&0.006&11.6&5.8&Keck\\
277.12971270&26.85534671&17.50&0.002&42.7&21.3&Spitzer\\
277.13008161&26.84907455&20.55&0.011&21.6&10.8&Spitzer,Keck\\
277.13061562&26.84941597&20.97&0.016&23.5&11.9&Spitzer,Keck\\
277.13116667&26.84513525&20.09&0.008&14.1&7.1&Keck\\
277.13117765&26.85026363&19.96&0.008&27.0&13.5&Spitzer,Keck\\
277.13143360&26.84232633&22.04&0.034&14.8&8.6&Keck\\
277.13152398&26.84961178&20.11&0.008&25.6&12.8&Spitzer,Keck\\
277.13173964&26.85489175&17.43&0.002&42.9&21.4&Spitzer\\
277.13310074&26.86055572&16.21&0.001&63.7&31.7&Spitzer,2MASS\\
277.13366038&26.82526749&16.28&0.002&69.6&34.7&Spitzer,2MASS\\
277.13461287&26.83484333&18.35&0.004&40.0&19.9&Spitzer\\
277.13763243&26.83161343&19.20&0.007&55.1&27.5&Spitzer\\
277.14074059&26.83242086&16.24&0.001&59.8&29.8&Spitzer,2MASS\\
277.14122122&26.84312777&17.25&0.002&45.4&22.6&Spitzer\\
277.14136220&26.84167755&17.19&0.002&46.4&23.1&Spitzer,2MASS\\
277.14168068&26.83999147&15.44&0.001&48.7&24.3&Spitzer,2MASS\\
277.14393913&26.84815732&16.63&0.001&56.4&28.1&Spitzer,2MASS\\
277.14408060&26.83259772&14.24&0.001&67.6&33.7&Spitzer,2MASS\\
277.14510022&26.83998378&17.04&0.002&59.4&29.6&Spitzer\\
277.14677186&26.83408275&14.78&0.001&72.1&35.9&Spitzer,2MASS\\
277.14732238&26.83496895&17.38&0.002&72.2&35.9&Spitzer\\
\enddata
\end{deluxetable}

\clearpage

\begin{deluxetable}{lrrrrrrr}
\centering
\tablecaption{Photometry for WISE 1828+2650  \label{phot}}
\tablehead{
\colhead{Telescope}&\colhead{Date,UT}&\colhead{Filter}&\colhead{$\lambda(\mu$m)}&\colhead{Mag. (mag)}&\colhead{SNR}&\colhead{F$_\nu(\mu$Jy)$^1$}}
\startdata
Keck/NIRC2 &2010-07-01&J$^3$&1.25&23.57$\pm$0.35&3&0.59$\pm$ 0.21\cr
HST/WFC3 &2011-05-09&F140W&1.39&23.15$\pm$0.09&11&0.75$\pm$ 0.07\cr
Keck/NIRC2 &Average&H&1.65&22.45$\pm$0.08$^2$&17&1.07$\pm$ 0.09\cr
Keck/NIRC2 &2010-07-01&H&1.65&22.52$\pm$0.23&5&1.34$\pm$ 0.25\cr
Keck/NIRC2 &2011-10-16&H&1.65&22.42$\pm$0.14&11&1.24$\pm$ 0.13\cr
Keck/NIRC2 &2012-07-09&H$^3$&1.65&22.46$\pm$0.12&11&1.40$\pm$ 0.12\cr
WISE &2010-03-30&W1$^4$&3.35&$>$18.5&&$<$12\cr
{\it Spitzer}/IRAC$^5$ &2010-07-10&Ch1&3.55&17.17$\pm$0.03&55&38.1$\pm$ 1.1\cr
{\it Spitzer}/IRAC &Average&Ch2&4.49&14.66$\pm$0.03&75&247$\pm$ 9\cr
{\it Spitzer}/IRAC &2010-07-10&Ch2&4.49&14.61$\pm$0.05&60&256$\pm$ 13\cr
{\it Spitzer}/IRAC &2010-12-04&Ch2&4.49&14.60$\pm$0.05&40&261$\pm$ 13\cr
{\it Spitzer}/IRAC &2011-11-29&Ch2&4.49&14.70$\pm$0.05&40&238$\pm$ 12\cr
{\it Spitzer}/IRAC &2012-05-25&Ch2&4.49&14.72$\pm$0.05&50&232$\pm$ 12\cr
WISE &2010-03-30&W2&4.6&14.39$\pm$0.06&20&299$\pm$ 18\cr
WISE &2010-03-30&W3$^4$&11.56&$>$12.5&&$<$290\cr
WISE &2010-03-30&W4$^4$&22.08&$>$8.8&&$<$2501\cr
\enddata
\tablecomments{$^1$Flux densities have {\it not} been color-corrected.$^2$Uncertainty weighted average, including 5\% calibration uncertainty.$^3$ Keck NIRC2/LGS result from \citet{Kirkpatrick2011}. $^4$ Upper limits are 95\% confidence. $^5$Includes  3\% IRAC calibration uncertainty.}
\end{deluxetable}

\begin{deluxetable}{lrrrrrr}
\centering
\tablecaption{Registration of Images \label{astrometry}}
\tablehead{\colhead{Tel-1/Tel-2}&\colhead{Date(UT)}&\colhead{Number of}&\colhead{$\Delta\alpha^1$}&
	\colhead{$\Delta\delta^1$}&\colhead{$\sigma_{mean}^{(2)}$}&\colhead{Rot. Unc.$^3$}\\
&&\colhead{Stars}&\colhead{(mas)}&\colhead{(mas)}&\colhead{(mas)}&\colhead{(deg)}}
\startdata
{\it HST}/2MASS$^4$	&2011-05-09	&11&10&-2	&$\pm$4&0.028\\ 
Keck/{\it HST}		&2010-07-01	&11&0.0&0.0	&$\pm$3&0.003\\
{\it Spitzer}/{\it HST}	&2010-07-10	&22&2&-11	&$\pm$11&0.010\\
{\it Spitzer}/{\it HST}	&2010-12-04	&21&17&11	&$\pm$16&0.017\\
Keck/{\it HST}		&2011-10-16	&21&1.7&0.0	&$\pm$3&0.003\\
{\it Spitzer}/{\it HST}	&2011-11-29	&16&-22&17	&$\pm$18&0.019\\
{\it Spitzer}/{\it HST}	&2012-05-25	&15&5&-16	&$\pm$13&0.015\\
Keck/{\it HST}		&2012-07-09	&15&-1.8&2.4	&$\pm$3&0.003\\
Keck/{\it HST}		&2012-10-07	&20&-0.2&-0.1	&$\pm$12&0.009\\
\enddata
\tablecomments{1) Difference between two reference frames; 2) Standard deviation of the mean, ($\sigma/\sqrt{N})$,in the offset in RA, DEC between stars observed with both telescopes; 3) Uncertainty in image rotation; 4) after removal of ($\Delta\alpha\, ,\Delta\delta$)(initial HST-2MASS)=(-0.50$\pm0.02$,0.15$\pm0.05$)$^{\prime\prime}$ offset.}
\end{deluxetable}

\begin{deluxetable}{lrrrrr}
\centering
\tablecaption{Astrometric Data for WISE 1828+2650  \label{astromdata}}
\tablehead{\colhead{Telescope}&\colhead{Date(UT)}&\colhead{MJD$^1$}&\colhead{RA (sec)$^2$}&
	\colhead{DEC(\arcsec)$^2$}&\colhead{Uncertainty(\arcsec)$^3$}}
\startdata
WISE$^4$&2010-Mar-30&55284.50&31.07805&37.9040&0.210\\
Keck1&2010-Jul-01&55378.44&31.10325&37.9443&0.029\\
Spitzer1&2010-Jul-10&55387.34&31.10961&37.9780&0.058\\
WISE$^4$&2010-Sep-28&55467.00&31.09463&37.8860&0.210\\
Spitzer2&2010-Dec-04&55534.27&31.12461&37.9020&0.062\\
Hubble&2011-May-09&55690.89&31.17538&38.0725&0.030\\
Keck2&2011-Oct-16&55850.21&31.19329&38.0785&0.025\\
Spitzer3&2011-Nov-29&55894.04&31.20339&38.1592&0.062\\
Spitzer4&2012-May-25&56072.25&31.24935&38.2111&0.061\\
Keck3&2012-Jul-08&56117.32&31.25462&38.3020&0.025\\
Keck4&2012-Oct-07&56207.22&31.25951&38.1087&0.040\\
\enddata
\tablecomments{1) MJD=Modified Julian Date; 2) Relative to $18^h28^m$ and $+26^o50^m$ (J2000.0); 3)Excludes $\sim$80 mas uncertainty in absolute 2MASS reference frame. 4) Not included in fitting due to large uncertainties.}
\end{deluxetable}

\begin{deluxetable}{lrrrr}
\centering
\tablecaption{Astrometric Data for WISE 1828+2650 With Respect to Source A \label{astromSrcA}}
\tablehead{\colhead{Telescope}&\colhead{MJD$^1$}&\colhead{$\Delta\alpha$(BD-SrcA,\arcsec)}&
	\colhead{$\Delta\delta$(BD-SrcA,\arcsec)}&\colhead{Uncertainty(\arcsec)$^2$}}
\startdata
Keck1&55378.44&4.647&3.042&0.032\\
Spitzer1&55387.34&4.609&3.179&0.121\\
Spitzer2&55534.27&4.896&3.009&0.122\\
HST&55690.89&5.570&3.192&0.018\\
Keck2&55850.21&5.853&3.175&0.015\\
Spitzer3&55894.04&5.884&3.264&0.134\\
Spitzer4&56072.25&6.535&3.301&0.134\\
Keck3&56117.32&6.615&3.374&0.016\\
Keck4&56207.22&6.695&3.303&0.042\\
\enddata
\tablecomments{1) MJD=Modified Julian Date; 2) Includes uncertainties due to extraction uncertainties, field rotation, residual Keck distortion.}
\end{deluxetable}

\begin{deluxetable}{lccc}
\centering
\tablecaption{Parallax And Proper Motion \label{parallax}}
\tablehead{\colhead{Parameter}&\colhead{Method-1$^1$}&\colhead{Method-2$^2$}&\colhead{Method-(1+2)$^{(2)}$}}
\startdata
$\mu_\alpha$ (\arcsec/yr)&0.969$\pm$0.017 & 0.936$\pm$0.015&0.954$\pm$0.011\\
$\mu_\delta$ (\arcsec/yr)&0.157$\pm$0.019 & 0.152$\pm$0.017&0.153$\pm$0.0125\\
Parallax, $\pi$,(\arcsec)&0.103$\pm$0.016 & 0.079$\pm$0.012&0.090$\pm$0.0095\\
Distance (pc)		&9.7$_{-1.1}^{+1.6}$& 12.6$_{-1.9}^{+2.2}$&11.2$_{-1.0}^{+1.3}$\\
$\alpha_{J2000.0}^3$ 	&$30.344^s\pm$0.077$^s$&$30.375^s\pm$0.012$^s$&$30.359s \pm0.009s$\\
$\delta_{J2000.0}^3$ &$36.120^{\prime\prime}\pm0.18^{\prime\prime}$&$36.200^{\prime\prime}\pm0.21^{\prime\prime}$&$36.278^{\prime\prime}\pm0.147^{\prime\prime}$\\
$\chi^2$ (with \# deg. of freedom)		&18.6\, (13)& 12.7\, (13)&36.3 (31) \\ \hline
Adopted Distance$^4$ (pc)	&\multispan{3}{11.2$_{-1.0}^{+1.3}$}\\
Tangential Velocity$^4$, $v_{tan}$ (km s$^{-1}$) &\multispan{3}{45$_{-10}^{+15}$}\\
\enddata
\tablecomments{1) Based on ``absolute coordinates"; 2) Based on offsets from Source A; 3) With respect to  $18^h28^m$ and $26^o50^\prime$. The random uncertainty does not include $\sim$80 mas uncertainty in absolute 2MASS reference frame. 4) From the combination of  Method-1 and Method-2. }
\end{deluxetable}

\begin{deluxetable}{lccccc}
\centering
\tablecaption{Cluster Kinematics and WISE 1828+2650 \label{UVW}}
\tablehead{\colhead{Name$^1$}&\colhead{Age}&\colhead{U$^{(2)}$ }&\colhead{V}&\colhead{W} &\colhead{Min $\Delta UVW^{(3)}$}\\
&(Myr)& (km s$^{-1}$)&(km s$^{-1}$)&(km s$^{-1}$)&(km s$^{-1}$)}
\startdata
Argus/IC2391&40&-22&-14.4&-5&46\\
TW  Hydrae&8&-11&-18&-5&50\\
Tucana/Horologium&30&-11&-21&0&55\\
$\beta$ Pictoris&12&-11&-16&-9&48\\
AB Doradus&50&-8&-27&-14&49\\
$\eta$ Chamaeleontis&8&-12&-19&-10&46\\
Columba&30&-13&-22&-6&50\\
\enddata
\tablecomments{$^1$Cluster data from \citet{Zuckerman2004b,Torres2008};$^2$Cluster velocities are calculated such that $U$ is positive toward the Galactic Center; $^3$Smallest difference between $(U,V,W)$ for WISE 1828 and a cluster within a range of unknown $V_{rad}$ between -100 and 100 km s$^{-1}$. }
\end{deluxetable}

\clearpage

\begin{deluxetable}{ccc|cc}
\tablecaption{ H Band Distortion Coefficients \label{XYmap}}
\tablehead{
\colhead{Parameter}&\colhead{X-Estimate}&\colhead{X-Uncertainty}&\colhead{Y-Estimate}&\colhead{Y-Uncertainty}}
\startdata
 const.	& -0.306229 & 6.16$\times 10^{-4}$ & -0.700962 & 6.16$\times 10^{-4}$ \\
 x 	& 0.00369728 & 2.44$\times 10^{-6}$& -0.00134633 & 2.44$\times 10^{-6}$ \\
 y 	& -0.0021500 & 2.39$\times 10^{-6}$ & -0.00333556 & 2.39$\times 10^{-6}$\\
 x$^2$ 	& 7.931$\times 10^{-7}$  & 1.17$\times 10^{-8}$ & 6.116$\times 10^{-6}$ & 1.17$\times 10^{-8}$\\
 x y 	& -1.112$\times 10^{-6}$ & 1.02$\times 10^{-8}$ & 2.258$\times 10^{-7}$ & 1.02$\times 10^{-8}$ \\
 y$^2$ 	& 1.0698$\times 10^{-5}$ & 1.15$\times 10^{-8}$ & 5.081$\times 10^{-6}$ & 1.15$\times 10^{-8}$\\
 x$^3$ 	& -5.857$\times 10^{-9}$ & 1.47$\times 10^{-11}$& -1.957$\times 10^{-10}$ & 1.47$\times 10^{-11}$ \\
 x$^2$y & 1.015$\times 10^{-9}$ & 1.39$\times 10^{-11}$ & 2.677$\times 10^{-9}$ & 1.39$\times 10^{-11}$ \\
 xy$^2$	& -3.869$\times 10^{-9}$ & 1.35$\times 10^{-11}$ & 1.060$\times 10^{-9}$ & 1.35$\times 10^{-11}$\\
 y$^3$ 	& 6.10$\times 10^{-10}$  & 1.36$\times 10^{-11}$ & 5.217$\times 10^{-9}$ & 1.36$\times 10^{-11}$ \\
 x$^4$ 	& -8.486$\times 10^{-12}$ & 5.21$\times 10^{-14}$& 9.835$\times 10^{-13}$ & 5.21$\times 10^{-14}$  \\
 x$^3$y & 9.712$\times 10^{-13}$ & 4.90$\times 10^{-14}$ & -3.382$\times 10^{-12}$ & 4.90$\times 10^{-14}$\\
 x$^2$y$^2$ & -5.878$\times 10^{-12}$ & 4.73$\times 10^{-14}$ & 2.080$\times 10^{-12}$ & 4.73$\times 10^{-14}$\\
 xy$^3$	& 2.695$\times 10^{-12}$ & 4.78$\times 10^{-14}$ & -4.396$\times 10^{-12}$ & 4.78$\times 10^{-14}$\\
 y$^4$	&-2.244$\times 10^{-12}$&4.88$\times 10^{-14}$& -2.219$\times 10^{-12}$&4.88$\times 10^{-14}$\\
\enddata
\end{deluxetable}

\clearpage

\clearpage

\clearpage

\clearpage

\clearpage

\clearpage

\clearpage

\clearpage

\clearpage

\end{document}